\documentclass[12pt]{article}
\catcode`\@=11
\def\numberbysection{\@addtoreset{equation}{section}
\def\theequation{\thesection.\arabic{equation}}}
\numberbysection
\newcommand{\abs}[1]{\vert#1\vert}
\newcommand{\beq}{\begin{equation}}
\newcommand{\beqa}{\begin{eqnarray}}
\newcommand{\bin}[2]{{#1\choose#2}}
\newcommand{\cond}[2]{P(#1\vert#2)}
\renewcommand{\d}{\mathrm{d}}
\newcommand{\db}[2]{$\matrix{#1\cr#2\cr}$}
\newcommand{\dbl}[2]{$\matrix{#1\hfill\cr#2\hfill\cr}$}
\renewcommand{\dim}[1]{\subsubsection*{$\bullet$ #1}}
\newcommand{\dsum}{\displaystyle\sum}
\newcommand{\eeq}{\end{equation}}
\newcommand{\eeqa}{\end{eqnarray}}
\newcommand{\eff}{_{\mathrm{eff}}}
\newcommand{\eps}{\varepsilon}
\newcommand{\eq}{_{\mathrm{eq}}}
\newcommand{\frad}[2]{\displaystyle{\displaystyle#1\over\displaystyle#2}}
\renewcommand{\max}{_{\mathrm{max}}}
\newcommand{\mean}[1]{\left\langle#1\right\rangle}
\newcommand{\prob}[1]{\mathrm{Prob}\big\{#1\big\}}
\renewcommand{\skew}{_{\mathrm{skew}}}
\newcommand{\tr}{_{\mathrm{tr}}}
\newcommand{\tripl}[3]{$\matrix{#1\cr#2\cr#3\cr}$}
\begin{document}
\centerline{\Large\bf Models of competitive learning:}
\vspace{.2cm}
\centerline{\Large\bf complex dynamics, intermittent conversions}
\vspace{.2cm}
\centerline{\Large\bf and oscillatory coarsening}
\vspace{1cm}
\centerline{\large
by Anita~Mehta$^{a,}$\footnote{Present and permanent address:
S.N.~Bose National Centre for Basic Sciences, Salt Lake,
Block JD, Sector III, Calcutta 700091,
India}$^,$\footnote{anita@boson.bose.res.in}
and J.M.~Luck$^{b,}$\footnote{luck@spht.saclay.cea.fr}
}
\vspace{1cm}
\centerline{$^a$Oxford Physics, Clarendon Laboratory, Parks Road,
Oxford~OX1~3PU, England}
\vspace{.3cm}
\centerline{$^b$Service de Physique Th\'eorique,
CEA Saclay, 91191 Gif-sur-Yvette cedex, France}
\vspace{1cm}
\begin{abstract}
We present two models of competitive learning, which
are respectively interfacial (i.e., where the interiors
of domains are unaffected by the rules of the model)
and cooperative (i.e., where the bulk as well as the interface
of an individual domain is governed by the rules of the model) learning.
This learning is outcome-related, so
that spatially and temporally local environments
influence the conversion of a given site between one of two different types.
We focus here on the behaviour of the
models at coexistence, which yields new critical
behaviour and the existence of a phase involving a novel
type of coarsening which is {\it oscillatory} in nature.
In the discussion we speculate on, among other issues, the likely behaviour
of the models away from coexistence.
\end{abstract}
\vfill
\noindent To appear in Physical Review E
\vskip -2pt
\noindent PACS: 02.50.Ey, 02.50.Le, 05.70.Ln, 05.40.+j, 64.60.Ht, 89.60.+x
\hfill T/99/086
\newpage
\section{Introduction}

The notion of an outcome is familiar in the social sciences, but
less so in the physical sciences.
It is used in the former to describe
the notion of a result following a course of action, especially in
the context of game-theoretic applications~\cite{L}.
In this paper we use this
concept to devise models which, while motivated by ideas in the
social sciences, are of interest to physicists both because
they lead to new phenomena, such as that of {\it oscillatory coarsening},
as well as because, via their connection with known physical models,
they are able to provide indications of new critical behaviour.

The motivation of our model comes from the phenomenon of competitive
learning;
imagine the existence of two species, distracted (D) and receptive (R).
We postulate that the D species is slower to learn
than the R species, but that, on the other hand, the R species is
more quick to forget; that is, a token R site
is more swayed by the short-term successes
of its neighbours to convert to their species.
This could be a model
of conservative versus radical societies, where the former adapt
relatively slowly to change, but are more stable once changed, whereas
the opposite applies to the latter.

We shall define and investigate successively an interfacial version
of our model in Sec.~\ref{s2}, and a cooperative version in Sec.~\ref{s3}.
The emphasis will be put on the phase diagram of the model at
coexistence, i.e., when neither of the species is favoured over the other;
we will there examine
the types of order present in the model,
and the universal characteristics of the transition points
between these phases.
In Sec.~\ref{s4}, we will discuss our results.

\section{Interfacial model}
\label{s2}

\subsection{Definition and generalities}

We here introduce the first of our two models (which we term the interfacial
model) and set up definitions and notations.
Individuals sit at the sites (or nodes) of a regular lattice
with coordination number $z$.
We shall most often consider the $d$-dimensional hypercubic lattice,
for which $z=2d$.
We represent the efficiency of the individual at site $i$
as a (time-dependent) Ising spin variable:
\beq
\eta_i(t)=\left\{\matrix{
+1&\mbox{if $i$ is of type R at time $t$},\hfill\cr
-1&\mbox{if $i$ is of type D at time $t$}.\hfill\cr
}\right.
\eeq

The efficiencies are dynamical variables, which are updated
according to the following dynamical rules.
\begin{itemize}
\item{{\bf Step 1: Majority rule for site types}

In this step, we update the efficiencies via a zero-temperature
(ferromagnetic) majority rule.
In other words, where a site is surrounded
by a majority of its own type, it stays the same, while if it is
surrounded by a majority of the other type, it changes.
In the absence of a majority, the site flips type with probability 1/2.

More formally, this rule consists of aligning the efficiency $\eta_i$
with the local field acting upon it, according to
\beq
\eta_i(t+\tau_1)=\left\{\matrix{
+1\hfill&&\mbox{if}\;h_i(t)>0,\hfill\cr
\pm1&\mbox{w.p.}\;1/2\hfill&\mbox{if}\;h_i(t)=0,\hfill\cr
-1\hfill&&\mbox{if}\;h_i(t)<0.\hfill\cr
}\right.
\label{step1}
\eeq
The local field,
\beq
h_i(t)=\sum_{j(i)}\eta_j(t),
\label{hloc}
\eeq
is defined as the sum of the efficiencies of the $z$ neighbouring sites $j$
of site $i$, and $\tau_1$ is the associated time step.
}
\item{{\bf Step 2: Association of outcomes with sites}

In this step, we assign to each site $i$ an outcome $\sigma_i$, which
also takes Ising-like values $+1$ and $-1$,
corresponding respectively to success or failure.
The (time-dependent) outcomes $\sigma_i$ are random but
correlated with the efficiencies $\eta_i$, namely
\beqa
&\mbox{if}\;\eta_i(t)=+1,\;\mbox{then}\;\sigma_i(t+\tau_2)=\left\{\matrix{
+1&\mbox{w.p.}&p_+,\hfill\cr
-1&\mbox{w.p.}&1-p_+,\hfill\cr
}\right.
\nonumber\\
&\mbox{if}\;\eta_i(t)=-1,\;\mbox{then}\;\sigma_i(t+\tau_2)=\left\{\matrix{
+1&\mbox{w.p.}&p_-,\hfill\cr
-1&\mbox{w.p.}&1-p_-,\hfill\cr
}\right.
\label{step2}
\eeqa
with $\tau_2$ being the associated time step.
The performance parameters $p_\pm$ characterise the strength of correlation
between efficiency and outcome for each species.

This step is meant to model the {\it performance} of
a particular site, according to its type.
Thus according to our rules, if $p_+>p_-$, the sites of type R are more
likely to succeed at learning some new process (thus
adapting more quickly to new patterns) than sites of type D.
}
\item{{\bf Step 3: Conversion of sites according to performances}

In this step, we model the `fickleness' of the sites
where, as mentioned above, we assume that the R sites
are more vulnerable to the performances of their neighbours
than are the D sites.

We denote by $N^+_i(t)$ the number of R-type neighbours of a given site $i$,
and by $N^-_i(t)=z-N^+_i(t)$ the number of its D-type neighbours.
We also denote by $I^+_i(t)$ [respectively, $I^-_i(t)$]
the number of these which have a positive outcome:
\beq
\matrix{
N^+_i(t)=\dsum_{j(i)}\frad{1+\eta_j(t)}{2}=\frad{z+h_i(t)}{2},\hfill&
I^+_i(t)=\dsum_{j(i)}\frad{1+\eta_j(t)}{2}\,\frad{1+\sigma_j(t)}{2},\hfill\cr
N^-_i(t)=\dsum_{j(i)}\frad{1-\eta_j(t)}{2}=\frad{z-h_i(t)}{2},\hfill&
I^-_i(t)=\dsum_{j(i)}\frad{1-\eta_j(t)}{2}\,\frad{1+\sigma_j(t)}{2}.\hfill\cr
}
\label{ni}
\eeq

Then we postulate that the efficiencies are updated according to:
\beqa
&\mbox{if}\;\eta_i(t)=+1\;\mbox{and}
\;\frad{I^+_i(t)}{N^+_i(t)}<\frad{I^-_i(t)}{N^-_i(t)},
\;\mbox{then}\;\eta_i(t+\tau_3)=\left\{\matrix{
-1&\mbox{w.p.}&\eps_+,\hfill\cr
+1&\mbox{w.p.}&1-\eps_+.\hfill\cr
}\right.
\nonumber\\
&\mbox{if}\;\eta_i(t)=-1\;\mbox{and}
\;\frad{I^-_i(t)}{N^-_i(t)}<\frad{I^+_i(t)}{N^+_i(t)},
\;\mbox{then}\;\eta_i(t+\tau_3)=\left\{\matrix{
+1&\mbox{w.p.}&\eps_-,\hfill\cr
-1&\mbox{w.p.}&1-\eps_-,\hfill\cr
}\right.
\label{step3}
\eeqa
with $\tau_3$ being the associated time step.
The efficiencies are not updated, i.e., they are left unchanged,
if none of the conditions given in eq.~(\ref{step3}) is satisfied.
The convertibility parameters $\eps_\pm$ characterise the fickleness
of each species.

The above rule is meant to hold only if the individual at site $i$
is able to compare the outcomes of both types of its neighbours,
namely for $N^+_i(t)$ and $N^-_i(t)\ne0$, i.e., $N^+_i(t)\ne0$ and $z$.
This is why we call this model the `interfacial' model,
since dynamical evolution occurs only for interfacial sites $i$
whose neighbours are thus {\it not} all of the same type.
Put another way, the interiors of clusters of any type do not evolve
according to its rules.

Then, if $p_+>p_-$ and $\eps_+>\eps_-$, we note that even though the R sites
are globally more successful than the D sites,
they are more susceptible to local performance-based conversions
than are the latter.
}
\end{itemize}

So far we have only defined three elementary steps
of the dynamics of the interfacial model,
whose time scales, $\tau_1$, $\tau_2$, $\tau_3$ are arbitrary.
We shall consider the realistic regime where the time scale over which
individuals change type, i.e., $\tau_1$ or $\tau_3$,
is much larger than the characteristic time scale of their activity
over which their outcomes are updated, i.e., $\tau_2$.
Throughout the following, we thus set for simplicity
\beq
\tau_2\to0,\qquad\tau_1=\tau_3=1,
\label{tau}
\eeq
and we consider an ordered sequential dynamics,
obtained by a regular `sweeping' of the system.

In the regime~(\ref{tau}) a given realisation of the outcomes
$\sigma_j(t)$
will enter the rules~(\ref{step3}) only once.
Hence steps 2 and 3 can be recast as effective dynamical rules
involving the efficiencies $\eta_i(t)$ and the associated
local fields $h_i(t)$ alone.
Obviously, these rules only hold for the sites $i$ which are interfacial at time $t$,
i.e., when $h_i(t)$ is different from its extremal values $\pm z$.
They are of the form:
\beqa
&\mbox{if}\;\eta_i(t)=+1,\;\mbox{then}\;\eta_i(t+1)=\left\{\matrix{
+1&\mbox{w.p.}&w_+(h_i(t)),\hfill\cr
-1&\mbox{w.p.}&1-w_+(h_i(t)),\hfill\cr
}\right.
\nonumber\\
&\mbox{if}\;\eta_i(t)=-1,\;\mbox{then}\;\eta_i(t+1)=\left\{\matrix{
+1&\mbox{w.p.}&w_-(h_i(t)),\hfill\cr
-1&\mbox{w.p.}&1-w_-(h_i(t)).\hfill\cr
}\right.
\label{w23}
\eeqa
The effective transition probabilities $w_\pm(h)$ can be evaluated explicitly,
by enumerating the $2^z$ possible realisations
of the outcomes $\sigma_j$ of the sites $j$ neighbouring site $i$,
and weighting them appropriately.
Consider for definiteness the example of the square lattice.
We have $z=4$, so that the extremal values of the local field are $h=\pm4$,
while its values at interfacial sites are $0$ and $\pm2$.
The corresponding transition probabilities read:
\beqa
w_+(+2)&=&1-\eps_+p_-(1-p_+^3),\nonumber\\
w_-(+2)&=&\eps_-(1-p_-)(1-(1-p_+)^3),\nonumber\\
w_+(0)&=&1-\eps_+p_-(1-p_+)(2-p_--2p_++3p_-p_+),\nonumber\\
w_-(0)&=&\eps_-p_+(1-p_-)(2-p_+-2p_-+3p_-p_+),\nonumber\\
w_+(-2)&=&1-\eps_+(1-p_+)(1-(1-p_-)^3),\nonumber\\
w_-(-2)&=&\eps_-p_+(1-p_-^3).
\eeqa

In order to predict the behaviour of the interfacial model
for generic values of the performance parameters $p_\pm$
and the convertibility parameters $\eps_\pm$,
it is advantageous to examine in most detail the symmetric situation
\beq
p=p_+=p_-,\quad\eps=\eps_+=\eps_-,
\label{sym}
\eeq
and then to investigate the influence of the two biasing fields
\beq
H=p_+-p_-,\quad B=\eps_--\eps_+.
\label{bias}
\eeq
Indeed in the symmetric case~(\ref{sym}), neither of the types R or D is favoured.
In other words, the rules of the symmetric model are invariant
under a global flip of all the spins.
The possible dynamical phase transitions between various kinds of collective
behaviour are therefore expected to take place in this symmetric situation,
in analogy with the phase transition in the ferromagnetic Ising model,
which takes place at zero magnetic field (symmetric situation)
and low enough temperature.
We will discuss at a qualitative level
the implications of this phase diagram for the generic situation
of non-zero biasing fields $B$ or $H$, in Sec.~\ref{s4}.

This symmetric situation models the case when the two species,
despite being identical in all their properties,
are nevertheless distinguishable as being of two distinct types.
The parameters $p$ and $\eps$ then control the intensity of surface noise.
We attach the label `surface' to them because, as mentioned above,
all discernible effects of these parameters are restricted
to the interfaces separating clusters of individuals
belonging to either of the two species.

In the rest of this section we shall explore the
behaviour of the symmetric interfacial model in the regime~(\ref{tau}),
via an approximate analytical treatment (known as the pair approximation)
and then by means of numerical simulations.
The basic quantities to be considered hereafter
are the magnetisation $M$ and the energy $E$.
These quantities are defined for a finite sample of $N$ individuals,
i.e., $N$ sites (or nodes) and $Nz/2$ bonds (or links), as
\beq
M=\frac{1}{N}\sum_i\eta_i,\qquad
E=\frac{1}{Nz}\sum_{(ij)}\left(1-\eta_i\eta_j\right).
\label{me}
\eeq
In the following we shall usually consider the mean values
$\mean{M}$ and $\mean{E}$, where the brackets represent an average
over the random initial configuration of efficiencies $\{\eta_i(0)\}$
and over the stochastic dynamical rules~(\ref{step1}),
(\ref{step2}), and~(\ref{step3}) --
i.e., the whole `thermal history' of the system.
The magnetisation $\mean{M}$ is the mean efficiency of an individual,
while the energy $\mean{E}$ is the proportion of `disparate' bonds $(ij)$,
such that $\eta_i\ne\eta_j$.
In the case of the cooperative model in three dimensions
(to be presented later),
we shall be led to consider also the dimensionless specific heat $C$.
In analogy with equilibrium systems,
this quantity is defined as the variance of the energy $E$ per bond, namely
\beq
C=\frac{Nz}{2}\left(\mean{E^2}-\mean{E}^2\right)
=\frac{1}{2Nz}\sum_{(ij)(k\ell)}
\Big(\mean{(1-\eta_i\eta_j)(1-\eta_k\eta_\ell)}
-\mean{1-\eta_i\eta_j}\mean{1-\eta_k\eta_\ell}\Big).
\label{defc}
\eeq

It will be advantageous to put the present model in perspective with the
kinetic Ising model investigated in refs.~\cite{O,DG}.
This is a dynamical model for Ising spins $\eta_i(t)$
on the square lattice,
defined by a stochastic rule of the form~(\ref{w23}), namely
\beq
\eta_i(t+1)=\left\{\matrix{
+1&\mbox{w.p.}&W(h_i(t)),\hfill\cr
-1&\mbox{w.p.}&1-W(h_i(t)).\hfill\cr
}\right.
\label{r12}
\eeq
The transition rates $W(h_i(t))$ depend only on the local field $h_i(t)$
defined in eq.~(\ref{hloc}).
They assume the most general form compatible with spin-flip symmetry,
i.e., $W(h)+W(-h)=1$.
This form involves two parameters $(0\le p_1, p_2\le 1)$:
\beqa
W(+4)&=&p_2,\nonumber\\
W(+2)&=&p_1,\nonumber\\
W(0)&=&1/2,\nonumber\\
W(-2)&=&1-p_1,\nonumber\\
W(-4)&=&1-p_2.
\label{w12}
\eeqa
On the ferromagnetic side $(1/2\le p_1, p_2\le1)$,
this model contains several special cases,
including the Ising model with Glauber dynamics,
the (noisy) voter model, and the majority vote model.
Introduced by de Oliveira~et~al.~\cite{O}
in the context of a general investigation of nonequilibrium spin models,
it was subsequently treated by Drouffe and Godr\`eche~\cite{DG}
who interpreted the parameters $p_1$ and $p_2$ as two effective temperatures,
linked respectively to an interfacial and a bulk noise.
As our symmetric model only contains interfacial noise,
it should correspond to the above model along the $p_2=1$ line,
with $1/2\le p_1\le1$ being a measure of the strength of interfacial noise.
On this line, the model is paramagnetic for a large enough surface noise
$(1/2\le p_1<3/4)$,
while it exhibits a frozen, i.e., totally ordered, ferromagnetic phase
for a weak enough noise $(3/4<p_1\le1)$.
The transition point $(p_1=3/4$, $p_2=1)$ corresponds to the voter model.
This model has been investigated extensively,
both by mathematicians~\cite{VM} and by physicists~\cite{VP}, and
is known to be critical in two dimensions.

\subsection{Analytical approach: pair approximation}
\label{s2a}

The pair approximation is a particular case of the cluster method,
an analytical approach proposed long ago~\cite{KI}
as a systematic improvement over mean-field theory
to provide approximate solutions of statistical mechanical models
(see ref.~\cite{BU} for a review).
Contrary to the standard mean-field approximation, or site approximation,
the pair approximation has the advantage of taking into account
correlations between pairs of neighbouring sites.
A dynamical variant of the pair approximation had been introduced
by Dickman~\cite{D} in the framework of a surface-reaction model.
More recently, de Oliveira~et~al. have also used a dynamical pair approximation
to investigate their two-parameter kinetic Ising model~\cite{O}.

We propose the following {\it fully dynamical pair approximation},
which leads to closed-form coupled evolution equations
for the mean magnetisation $M(t)$ and energy $E(t)$
in our symmetric interfacial model; in the preceding,
as well as in what follows,
the angular brackets implying averaging
have been taken to be implicit.

The basic object of the pair approximation is the configuration
of the dynamical variables attached to a bond.
In the present case of a spatially homogeneous system,
with binary site variables $\eta_i$,
an (oriented) bond $(ij)$ can assume four configurations.
We introduce the corresponding probabilities
\beqa
P(++)&=&\prob{\eta_i=+1\;\mbox{and}\;\eta_j=+1}=x,\nonumber\\
P(--)&=&\prob{\eta_i=-1\;\mbox{and}\;\eta_j=-1}=y,\nonumber\\
P(+-)&=&\prob{\eta_i=+1\;\mbox{and}\;\eta_j=-1}=\frad{1-x-y}{2},\nonumber\\
P(-+)&=&\prob{\eta_i=-1\;\mbox{and}\;\eta_j=+1}=\frad{1-x-y}{2}.
\eeqa

The pair approximation consists of reducing any observable,
i.e., the expectation of any function of the efficiencies $\eta_i$,
to a function of $x$ and $y$, by systematically neglecting
dynamical correlations between
the efficiencies of any two sites which are not first neighbours,
as well as any higher-order correlations.

For instance, the probability law of the efficiency of any given site reads:
\beqa
P(+)&=&\prob{\eta_i=+1}=P(++)+P(+-)=\frac{1+x-y}{2},\nonumber\\
P(-)&=&\prob{\eta_i=-1}=P(-+)+P(--)=\frac{1-x+y}{2}.
\eeqa
The mean magnetisation $M$ and energy $E$ hence read:
\beq
M=x-y,\quad E=1-x-y,\qquad\mbox{i.e.,}\quad
x=\frac{1-E+M}{2},\quad y=\frac{1-E-M}{2}.
\eeq
Thus, knowing $x$ and $y$ is equivalent to knowing $M$ and $E$.

As a second example,
consider the conditional probability $\cond{\sigma}{\tau}$,
defined as being the probability that $\eta_i=\sigma$,
given that $\eta_j=\tau$ for one of the neighbours $j$ of site $i$.
We have:
\beq
\matrix{
\cond{+}{+}=\frad{P(++)}{P(+)}=\frad{2x}{1+x-y},\hfill&&
\cond{-}{+}=\frad{P(-+)}{P(+)}=\frad{1-x-y}{1+x-y},\hfill\cr\cr
\cond{+}{-}=\frad{P(+-)}{P(-)}=\frad{1-x-y}{1-x+y},\hfill&&
\cond{-}{-}=\frad{P(--)}{P(-)}=\frad{2y}{1-x+y}.\hfill
}
\label{cond}
\eeq

The standard mean-field approximation corresponds to setting
$\cond{+}{\sigma}=P(+)=(1+M)/2$ and $\cond{-}{\sigma}=P(-)=(1-M)/2$
for $\sigma=\pm1$.
The variables $x$ and $y$ are then related by $(x-y)^2-2(x+y)+1=0$,
i.e., $E=(1-M^2)/2$.

Within the dynamical pair approximation,
coupled evolution equations for $x(t)$ and $y(t)$,
or equivalently, for $M(t)$ and $E(t)$,
can be derived by enumerating all the configurations of efficiencies
attached to a given bond and its neighbourhood,
weighting each of them with the appropriate probability,
and by the appropriate transition rate for each step of the dynamics,
as given in eqs.~(\ref{step1}), (\ref{step2}), and~(\ref{step3}).

We shall give just one example of how our calculations go.
Consider a $++$ bond $(ij)$,
and apply step 1 of the dynamics to site $i$ at time $t$, say.
The rule~(\ref{step1}) involves the local field $h_i(t)$,
or equivalently $N^+_i(t)=(z+h_i(t))/2$, defined in eq.~(\ref{ni}).
Consider the $z$ neighbours of site $i$.
We know that the neighbour $j$ is such that $\eta_j(t)=+1$.
The efficiencies $\eta_k$ of the $(z-1)$ other neighbours $k$
are treated as independent random variables,
each of them being distributed according to the conditional
probabilities $\cond{\eta_k}{+}$.
The random variable $N^+_i(t)$
is thus distributed according to the binomial law
\beqa
\prob{N^+_i=n}&=&\bin{z-1}{n-1}\cond{+}{+}^{n-1}\cond{-}{+}^{z-n}\nonumber\\
&=&\bin{z-1}{n-1}\frac{(2x)^{n-1}(1-x-y)^{z-n}}{(1+x-y)^{z-1}}
\qquad(n=1,\dots,z).
\eeqa
The probabilities of occurrence of the various cases of the
rule~(\ref{step1}) are thus known.

This procedure leads to coupled first-order differential equations, of the form
\beq
\frac{\d M}{\d t}
=\left(\frac{\d M}{\d t}\right)_1+\left(\frac{\d M}{\d t}\right)_3,
\qquad \frac{\d E}{\d t}
=\left(\frac{\d E}{\d t}\right)_1+\left(\frac{\d E}{\d t}\right)_3,
\label{i}
\eeq
where the contributions of the various steps of the dynamics appear
additively.
The combinatorial analysis involved in the expressions of these contributions
depends strongly on the lattice under consideration.

\dim{Two-dimensional case}

In the two-dimensional case, on the square lattice,
the various contributions to the differential equations~(\ref{i}) read:
\beqa
\left(\frac{\d M}{\d t}\right)_1
&=&\frac{2E^2M}{(1-M^2)^2}\bigg(3(1-M^2)-4E\bigg),\nonumber\\
\left(\frac{\d M}{\d t}\right)_3
&=&\frac{4\eps p(1-p)E^2M}{(1-M^2)^3}
\bigg((5p^2-5p-2)(3+M^2)E^2-4p(5p-9)(1-M^2)E\nonumber\\
&&{\hskip 3cm}+3p(p-5)(1-M^2)^2\bigg),\nonumber\\
\left(\frac{\d E}{\d t}\right)_1
&=&\frac{2E^3}{(1-M^2)^3}\bigg(\!-2(1-M^4)+3(1+M^2)E\bigg),\nonumber\\
\left(\frac{\d E}{\d t}\right)_3
&=&\frac{4\eps p(1-p)E}{(1-M^2)^3}
\bigg(\!-2(2p-1)(1+3M^2)E^3+2p(p+3)(1-M^4)E^2\nonumber\\
&&{\hskip 2.5cm}-3(p^2+p+1)(1-M^2)^2E+(p^2+p+1)(1-M^2)^3\bigg).\nonumber\\
\label{i2}
\eeqa

Although these equations cannot be integrated explicitly,
the dynamical phase diagram within the pair approximation
can be extracted from them as follows.
The steady-state values of the magnetisation and of the energy
are determined by equating the right-hand side of eqs.~(\ref{i2}) to zero.
The criterion for the presence of collective behaviour, i.e., long-range order,
is the instability of the fixed-point solution such that $M=0$.

A linear stability analysis thus allows us to predict the existence
of a transition line in the $p$-$\eps$ plane, given by
\beq
\eps_c(p)=\frac{5}{4p(1-p)(1+7p+3p^2)}.
\label{pa2}
\eeq
This line has two endpoints, $p_{c1}=0.47598$ and $p_{c2}=0.83060$,
corresponding to transitions as a function of $p$, if we set
$\eps=1$.
The intermediate phase $(p_1<p<p_2$ and $\eps_c(p)<\eps\le1)$
is paramagnetic, i.e., disordered,
while the rest of the parameter space is frozen,
i.e., totally ferromagnetically ordered.

\dim{Three-dimensional case}

In the three-dimensional case, on the cubic lattice,
the coupled differential equations for $M(t)$ and $E(t)$
are much more lengthy than eqs.~(\ref{i2}).
We prefer not to write them down in full.
Within this approximation, we predict the same phase diagram
as for the square lattice, with
\beq
\eps_c(p)=\frac{292}{3p(1-p)(32+1217p-2243p^2+2437p^3-245p^4)}.
\label{pa3}
\eeq
This line again has two endpoints, $p_{c1}=0.53047$ and $p_{c2}=0.86908$,
if we set $\eps=1$ as before.
We contrast the above with the predictions of standard mean-field theory
which does {\it not} predict a phase transition for the symmetric
interfacial model, in either two or three dimensions.

The qualitative phase diagram of the interfacial model is shown in Fig.~1,
together with that of the cooperative model investigated later on.
Quantitative data concerning the location and the nature
of the corresponding phase transitions are provided in Table~1.

\subsection{Numerical results}
\label{s2n}

We have performed a numerical investigation of the interfacial model
throughout the $p$-$\eps$ plane, both on the square and on the cubic lattice.
The quantitative measurements have been performed for $\eps=1$,
where the outcomes play a maximal role in the ordering behaviour of the
model via step 3.
The limiting situations $p=0$ and $p=1$ both represent deterministic outcomes,
where both species are restricted respectively
to being total failures and total successes;
in this case steps 2 and 3 are rendered essentially irrelevant,
and the majority rule in step 1 lets the ordering
proceed simply according to the types of the sites.

We therefore discuss in the following
the interfacial model on the square and on the cubic lattice,
for $\eps=1$, as a function of $p$.
In both cases we see clear evidence of a disordered,
paramagnetic phase for $p_{c1}<p<p_{c2}$, between two frozen phases.
This global picture is in qualitative agreement with the predictions
of the pair approximation described in Sec.~\ref{s2a}.

\dim{Two-dimensional case}

We have already underlined the analogy between our interfacial model
on the square lattice
and the two-parameter model investigated in refs.~\cite{O,DG}
along the $p_1=1$ line, corresponding to interfacial noise alone.
The strength of noise is measured by $p_2$ in that model,
while it is measured by $p$ and $\eps$ in our model.
This suggests that the phase transitions at $p_{c1}$ and $p_{c2}$,
which separate the disordered phase from both frozen phases,
are critical points which belong to the universality class of the voter
model: the intermediate disordered phase $(p_{c1}<p<p_{c2})$
corresponds to $p_2<3/4$,
while the two frozen phases $(0<p<p_{c1}$ and $p_{c2}<p<1)$
correspond to $p_2>3/4$.

Fig.~2 shows snapshots of the dynamics
of the model at times $t=8$, $t=64$, and $t=512$,
with a random initial condition.
In anticipation of our numerical result $p_{c2}\approx 0.7$ (see below),
we have chosen a value of the probability $p=0.7$ so that our snapshots
represent the time evolution of our system at (or very near)
the critical point.
Just as in ref.~\cite{DG}, the plots show a portion (of size~$256^2$)
of a square sample (of size~$512^2$) with periodic boundary conditions.
These plots bear a strong resemblance to those corresponding
to the voter model (see Fig.~5 of ref.~\cite{DG}),
thus confirming our expectations.

In order to provide quantitative confirmations of these observations,
we have studied the decay of the mean energy $E(t)$,
starting with a random initial configuration.
The two-dimensional voter model
is known to exhibit unusual critical behaviour~\cite{VM,VP};
this phenomenon is closely linked to the fact that
$d=2$ is the marginal dimensionality for Brownian motion,
which is known to be recurrent for $d<2$ and transient for $d>2$.
In particular, the mean energy $E(t)$ of the two-dimensional voter model
falls off very slowly, as
\beq
E(t)\approx\frac{\pi/2}{\ln(t/t_0)},
\label{ev}
\eeq
where the numerator $\pi/2$ is universal
(i.e., it is independent of the microscopic details of the dynamics),
whereas the time scale $t_0$ is not.

We have plotted in Fig.~3 the inverse energy $1/E(t)$ against $\ln t$,
for various values of $p$ in the vicinity of the phase transition
at $p_{c2}\approx 0.70$, gleaned from our snapshots of the system.
Each curve is an average over 200 independent samples of size~$500^2$.
The graphs curve downwards in the disordered phase,
where the energy converges to a non-zero equilibrium value $E\eq(p)$,
while they curve upwards in the frozen phases,
where the energy falls off as $E(t)\sim t^{-1/2}$;
the latter is because of an underlying diffusive type of
coarsening behaviour~\cite{AB}.
For $p=0.70$, the data seem to become asymptotically parallel
to the dashed line with slope $2/\pi$ [see eq.~(\ref{ev})], shown on the plot.
This observation confirms that our interfacial model
belongs to the universality class of the voter model
in a strong sense, i.e., including the prefactor of the law~(\ref{ev}).
Moreover, we thus obtain a rather accurate estimate for the transition
point: $p_{c2}=0.70\pm0.01$.
A similar analysis yields $p_{c1}=0.56\pm0.01$.
The numerical values of both transition points are listed in Table~1.

\dim{Three-dimensional case}

In three dimensions, on the cubic lattice,
we again find evidence of an intermediate disordered phase
for $p_{c1}<p<p_{c2}$.
Contrary to the two-dimensional case,
we do not have any a priori knowledge
of the phase transitions at $p_{c1}$ and $p_{c2}$.
In fact we do not know of {\it any} three-dimensional critical phenomenon
driven by interfacial noise alone.
It is nevertheless to be expected that the critical behaviour
at these phase transitions corresponds to a generic fixed-point behaviour,
with finite, non-trivial values of the various critical exponents.
This is in sharp contrast to the two-dimensional case,
which belongs to the very special universality class of the voter model,
with its well-known logarithmic behaviour~(\ref{ev}).
In particular, we expect a non-trivial power-law decay
of the mean energy at the transition points, of the form
\beq
E(t)\sim t^{-\Omega}.
\label{ep}
\eeq

This expectation is corroborated by our numerical results.
Fig.~4 shows a log-log plot of the energy $E(t)$,
for various values of $p$ in the vicinity of $p_{c1}$.
Each curve is an average over 50 independent samples of size~$100^3$.
The two extremal dashed lines show the expected behaviour
in the frozen and disordered phases,
while the intermediate one corresponds to the critical law~(\ref{ep}),
with an exponent $\Omega\approx0.1$.
This picture is confirmed by a more refined analysis
whose results are plotted in Fig.~5.
This graph shows a log-log plot of the effective exponent $\Omega\eff(t)$
(defined as being the negative of the slope of the least-square fit to
the data shown in Fig.~4, over the range $t/2<t'<2t$).
The different kinds of behaviour corresponding to the frozen and disordered
phases and to the critical point, appear more clearly.
The data for $p=0.40$, 0.42, and possibly 0.44, curve upwards
and eventually tend toward the value $1/2$, characteristic of the frozen phase.
Conversely, the data for $p=0.48$ and 0.50 curve downwards
and eventually tend toward the value $0$,
characteristic of the disordered phase.
The data for $p=0.45$ and 0.46 seem to converge to a non-trivial
critical exponent $\Omega$.
We thus obtain the estimates $p_{c1}=0.45\pm0.01$ and
$\Omega=0.10\pm0.04$.
A similar analysis leads to $p_{c2}=0.865\pm0.005$,
with an exponent $\Omega$ compatible with the above value.
Although our numerical data do not allow us to rule out
logarithmic behaviour of the mean energy,
they definitely point towards critical behaviour
characterised by a power law of the form~(\ref{ep}),
with a finite exponent $\Omega$.
The numerical values of the transition points and of the exponent
$\Omega$ are listed in Table~1.

\section{Cooperative model}
\label{s3}

\subsection{Definition}

One of the most interesting features of the model that
we have investigated in Sec.~\ref{s2} is its totally interfacial
behaviour.
However, from the viewpoint of the sociological behaviour that
we are trying to model, it lacks an important feature to do
with learning from the failures of neighbours when
all the sites concerned are of the {\it same} type.
Thus, consider a site of type D surrounded by others
of its own kind in the region of phase space where $p_+>p_-$:
it is then plausible that a majority of the surrounding sites
could fail at any given iteration.
However, according to the rules of the interfacial model
defined in the preceding section, the effect of this would be ephemeral;
in step 3 the central site would be converted to a site of type R,
but at the very next time step, the majority rule of step 1 would ensure
that it was converted back to a D site.
Thus there would be no long-term learning of the central site
from the failures of sites of its own kind,
to mirror the sociological phenomenon of
`learning from one's own mistakes'.

To incorporate such long-term learning, we introduce a modification
to our earlier model.
In essence this ensures that
the learning from failure in step 3 has a cooperative aspect,
that is the central site {\it as well as} its neighbours
learn from their failures and convert collectively to the other species.
This in turn ensures that the majority rule of step 1
does not interfere with the conversion based on learning in step 3,
since the central site and its neighbours are now all of the same kind,
i.e., R, in the case of the above example.
This rule (hard cooperative rule) is embodied in step 3a below.

A modification of the above would be to say that
only those sites which had failed, as well as the central site,
would convert to the other species:
then if a majority of surrounding sites fail,
those sites as well as the central sites would
convert at the next iteration to the other type.
This rule (soft cooperative rule) is embodied in step 3b below.
It is indeed a `softer' version of step 3a;
an alternative way of framing such a rule would be to have a stochastic
formulation where a noise would control the conversion of sites in step~3,
in proportion to the original number of failures.

In contrast with the interfacial model studied so far,
these variants of our cooperative model
incorporate both surface noise and bulk noise.
In this respect, the cooperative model is a generic dynamical model:
in two dimensions, and at a qualitative level, it can be mapped onto the
phase diagram of the two-parameter kinetic Ising model described above.
It will turn out, however, that our cooperative model possesses a much
greater diversity
of behaviour than the two-parameter model, since in addition to
the conventional ferromagnetic and paramagnetic phases,
it manifests a novel phase where the coarsening is {\it oscillatory}.

In the following we focus on step 3a.
We speculate that the soft version of step 3b
would only lead to a crossover behaviour from the pure interfacial model
of Sec.~2, to the hard cooperative model defined by steps 1 to 3a.
The rules discussed above are formulated respectively by:
\begin{itemize}
\item{{\bf Step 3a:
Cooperative conversion of sites according to performance (hard)}

We add to step 3 by considering the case when a site $i$
is surrounded by neighbours which are all of the same kind,
either all R, i.e., $N^+_i(t)=z$,
or all D, i.e., $N^-_i(t)=z$, with the notation of eq.~(\ref{ni}).
Then we postulate that:
\beqa
& &\mbox{if}\;N^+_i(t)=z\;\mbox{and}\;I^+_i(t)<z/2,
\;\mbox{then}\;\eta_i(t+\tau_3)=\eta_j(t+\tau_3)=-1,\nonumber\\
& &\mbox{if}\;N^-_i(t)=z\;\mbox{and}\;I^-_i(t)<z/2,
\;\mbox{then}\;\eta_i(t+\tau_3)=\eta_j(t+\tau_3)=+1,
\label{step3a}
\eeqa
where $j$ denotes all the $z$ neighbours of the central site $i$.
This step of hard cooperative conversion always involves the maximal
number $z+1$ of individuals in a cluster made up of any central site
and its neighbours.
}
\item{{\bf Step 3b:
Cooperative conversion of sites according to performance (soft)}

In this case we postulate that the cooperative rearrangement
described by the rule~(\ref{step3a}) only applies to the central site $i$,
and to the $z-I^\pm_i(t)$ neighbours $j$ of site $i$ who failed at time $t$.
Hence this step of soft cooperative conversion
involves a variable number $z+1-I^\pm_i(t)$ of individuals.
}
\end{itemize}

In the following, we analyse this cooperative model with the hard rule
(step 3a), along previous lines,
and find that although it is more generic in some respects,
in other ways it exhibits some very novel behaviour
to do with the {\it intermittent conversions} of sites.

We notice that, in the regime defined in eq.~(\ref{tau}),
steps 2 and 3a can be recast as effective dynamical rules
involving the efficiencies $\eta_i(t)$ alone, in analogy with eq.~(\ref{w23}).
These rules are valid only for sites $i$ which are bulk sites at time $t$
(i.e., when $h_i(t)=\pm z$), and are of the form
\beq
\matrix{
\mbox{if}\;\;h_i(t)=+z,\hfill&\mbox{then}\;\eta_i(t+1)=\eta_j(t+1)=-1
\hfill&\mbox{w.p.}\;\;\Pi,\cr
\mbox{if}\;\;h_i(t)=-z,\hfill&\mbox{then}\;\eta_i(t+1)=\eta_j(t+1)=+1
\hfill&\mbox{w.p.}\;\;\Pi,
}
\label{w23a}
\eeq
where it is understood that the efficiencies are not updated
in the complementary cases.
The probability $\Pi$ can be evaluated
by enumerating the $2^z$ possible realisations
of the outcomes $\sigma_j$ and weighting them appropriately.
We thus obtain
\beq
\Pi=\left\{\matrix{
(1-p)^3(1+3p)\hfill&\mbox{for}\;d=2,\cr
(1-p)^4(1+4p+10p^2)&\mbox{for}\;d=3.
}\right.
\label{pidef}
\eeq

\subsection{Analytical approach: pair approximation}
\label{s3a}

In order to analyse the cooperative model with the hard rule~(\ref{step3a})
within the pair approximation,
we have to determine the extra contribution of step 3a of the dynamics
to the differential equations~(\ref{i}).
The combinatorial analysis involved again depends on the lattice under
consideration.

\dim{Two-dimensional case}

In the case of the square lattice, the analysis is as follows.
Consider for concreteness the first line of eq.~(\ref{step3a}).
If the central spin is $\eta_i=+1$,
the move takes place with probability $\Pi\,p(+)\cond{+}{+}^4$;
it changes the total magnetisation by $N\Delta M=-10$,
while it leaves the energy unchanged.
If the central spin is $\eta_i=-1$,
the move takes place with probability $\Pi\,p(-)\cond{+}{-}^4$;
it changes the total magnetisation by $N\Delta M=-8$
and the total energy by $2N\Delta E=-4$.

We thus obtain after some algebra
\beqa
\left(\frac{\d M}{\d t}\right)_{3a}
&=&\frac{2(1-p)^3(3p+1)M}{(1-M^2)^3}
\bigg((3+M^2)E^4-40(1-M^2)E^3\nonumber\\
&&{\hskip 3.8cm}+30(1-M^2)^2E^2-5(1-M^2)^3\bigg),\nonumber\\
\left(\frac{\d E}{\d t}\right)_{3a}
&=&-\frac{2(1-p)^3(3p+1)E^4(1+3M^2)}{(1-M^2)^3}.
\label{i3a}
\eeqa
A linear stability analysis of the disordered solution $M=0$
of eqs.~(\ref{i}), including the contribution~(\ref{i3a}),
leads to the following prediction.
Within the pair approximation,
the model has a unique transition at $p_c=0.86805$,
between a disordered phase at $p<p_c$
and a ferromagnetically ordered one at $p>p_c$.
This prediction for the transition point is listed in Table~1.

\dim{Three-dimensional case}

The case of the cubic lattice is very similar to that of the square lattice.
The pair approximation again predicts the existence of a unique
phase transition, at $p_c=0.87319$.

\subsection{Numerical results}
\label{s3n}

For the cooperative model, updated in an ordered sequential way with rule 3a,
either on the square lattice or on the cubic lattice,
we see clear evidence of an intermediate disordered,
paramagnetic phase for $p_o<p<p_c$ (Fig~1).
The phase for $p_c<p<1$ is ferromagnetically ordered,
while the phase for $0<p<p_o$ {\it exhibits a novel kind of dynamical order}
that we call `oscillatory'.
These observations are illustrated in the two-dimensional case
in Figs.~6 and~7,
which show snapshots of the model for $p=0.1$ (oscillatory phase)
and $p=0.9$ (ferromagnetic phase), respectively.

\subsubsection{Ferromagnetic phase transition}

The phase transition at $p=p_c$ might be expected to be in the
universality class of the ferromagnetic Ising model,
since it demarcates a disordered phase from a ferromagnetically ordered one.
We have investigated this ferromagnetic phase transition
in two and three dimensions, by means of numerical simulations.

\dim{Two-dimensional case}

On the square lattice, the situation is very clearly seen from
Fig.~8, which shows a plot of the mean magnetisation $\mean{M}$ against $p$.
The samples used, of sizes~$100^2$ and~$200^2$, were large enough
to prevent the system from flipping between one ordered phase
and another during the simulations.
We have also checked that the data shown in Fig.~8 are not affected by
finite-size effects;
indeed the data for the two system sizes mentioned above cannot be
distinguished from one another on the plot.
The dashed line shows a fit of the form $\mean{M}\approx A(p-p_c)^{1/8}$,
with the well-known magnetisation exponent $\beta=1/8$
of the two-dimensional Ising model.
The quality of the plotted fit confirms our expectation:
the ferromagnetic transition we observe does indeed belong to the universality
class of the ferromagnetic Ising model.
We also obtain an accurate evaluation of the transition point,
$p_c=0.873\pm0.002$, i.e., $\Pi_c=(7.4\pm0.3)10^{-3}$, which is listed in
Table~1.

\dim{Three-dimensional case}

On the cubic lattice,
the situation is less clear cut as a consequence of huge finite-size effects.
An analysis of the mean magnetisation, analogous to that of Fig.~8,
is totally inefficient, even for the purpose of locating the transition.
We have instead considered the specific heat $C$, defined in eq.~(\ref{defc}).
Fig.~9 shows a plot of the specific heat $C$
(multiplied by the transition probability $\Pi$), against $p$,
for cubic samples of size $L^3$, with various values of $L$.
Each data point corresponds to an average over at least 100,000 time steps.
It is natural to consider the product $\Pi\,C$, instead of $C$ alone,
because the specific heat is due to bulk fluctuations,
whose driving force is the cooperative conversion embodied in step 3a;
these take place, as mentioned before, at each time step with probability
$\Pi$ (cf.~eq.~(\ref{pidef})).
We anticipate that the transition probability $\Pi$
will be very small throughout the ferromagnetic phase,
just as in the two-dimensional case.

The data for $\Pi\,C$ exhibit a peak, whose position, height and width
depend on $L$.
The critical exponents $\alpha$ and $\nu$ can in principle
be extracted from these data
by means of the finite-size scaling law~\cite{FSS}
\beq
C\approx L^{\alpha/\nu}F\left((p-p_c)L^{1/\nu}\right),
\eeq
which implies that the height of the peak scales as $L^{\alpha/\nu}$,
while its relative position $p\max(L)-p_c$
with respect to the genuine transition point as well
as its width, both scale as
$L^{-1/\nu}$.
Fig.~10 shows a plot of the peak position $p\max(L)$ against $L^{-1/\nu}$.
The fit shown as a dashed line incorporates a linear correction to scaling,
of the form $p\max(L)=p_c+L^{-1/\nu}(A+B\,L^{-\omega})$,
and uses the recent estimates $\nu=0.630$ and
$\omega=0.80$~\cite{GUI} corresponding to the universality class
of the three-dimensional Ising model.
Reasonable agreement is found,
although the amplitude for corrections to scaling is unexpectedly large.
We thus obtain the estimate $p_c=0.82\pm0.01$
for the critical ferromagnetic transition, i.e., $\Pi_c=(11\pm2)10^{-3}$.
We conclude therefore that the transition is most probably weakly Ising-like,
by which we mean that corrections to scaling are very large,
or else that the critical region is anomalously small.

\subsubsection{Oscillatory phase transition}

For small values of the parameter $p$,
i.e., values of the probability $\Pi$ close to unity,
the cooperative system exhibits an oscillatory phase,
already mentioned in the context of Fig.~7.

This phase in unusual in several respects, and we
were led to formulate its phenomenology as a result
of various numerical investigations.
First, since $\Pi$ is close to unity, the net outcome of every time step
is approximately to change all the efficiencies into their `opposites'.
This feature of the microscopic dynamics will be ignored
hereafter
by either only looking at the system at even times,
or by considering quantities such as the skew magnetisation
\beq
M\skew(t)=(-1)^tM(t).
\eeq
Second, and more importantly, we see that after a transient period,
the system is filled with clusters which keep evolving forever.
This perpetual motion is illustrated in Fig.~11,
which shows a plot of the evolution of the skew magnetisation against $t$,
for a square sample of size~$100^2$, with $p=0.1$ (hence $\Pi=0.990711$).
The skew magnetisation exhibits slow, irregular oscillations,
with a rather well-defined amplitude (roughly independent of size)
and period (scaling roughly as $L$).
This collective oscillatory behaviour remains noisy in the $p\to0$ limit,
because of the presence of `floppy' interfacial sites,
whose local field vanishes.
In order to investigate the transition at $p=p_o$ between this unusual
oscillatory phase and the usual disordered (paramagnetic) phase,
we have used the order parameter $\mean{M^2}=\mean{M\skew^2}$,
which is clearly non-zero in the oscillatory phase.

\dim{Two-dimensional case}

Fig.~12 shows a plot of the order parameter $\mean{M^2}$ against $p$,
for square samples of various sizes.
Each data point corresponds to an average over at least 100,000 time steps.
The curves clearly exhibit a common point,
corresponding to the transition.
In the oscillatory phase, the order parameter has a non-zero
thermodynamical value $\mean{M^2}_o\approx 0.16$,
with little or no dependence on $p$.
It is discontinuous at the transition point,
equaling $\mean{M^2}\tr\approx 0.06$ at the transition point itself,
and vanishing in the disordered phase.
In the vicinity of the transition point, the order parameter is found
to obey a finite-size scaling law of the form
\beq
\mean{M^2}\approx \Phi\left((p-p_o)L^{1/\nu_o}\right),
\label{mfss}
\eeq
illustrated in Fig.~13.
The best data collapse is obtained for $p_o=0.136\pm0.003$
and $1/\nu_o=0.45\pm0.10$.

\dim{Three-dimensional case}

On the cubic lattice, the situation is qualitatively similar.
The order parameter is still discontinuous,
with a thermodynamical value around
$\mean{M^2}_o\approx 0.16$ in the oscillatory phase.
The values of the transition point, $p_o=0.42\pm0.005$,
of the associated exponent characterising the size dependence,
$1/\nu_o=0.8\pm0.2$, and of the order parameter at the transition point,
$\mean{M^2}\tr\approx 0.0017$ (with a large error bar),
have been read off from Fig.~14.
Each data point again corresponds to an average
over at least 100,000 time steps.

\section{Discussion}
\label{s4}

Starting from simple and intuitive ideas about the nature of
experiential learning,
we have devised two models involving competitive dynamics.
The ideas are based on everyday experience:
individuals who are quick to learn are often those
who are also quick to adapt to changing circumstances (R type),
while those who are slow to learn (D type)
tend to be more conservative about what they know,
adapting relatively slowly to changes in their environment.
Our models embody these ideas, and emphasise the fact
that adaptivity may occasionally lead to an overly quick change of behaviour
(known colloquially as `knee-jerk reactions'),
and that in this sense slow and steady learners
can occasionally get the better of their quick counterparts.

Both models have the same essential features:
first, a zero-temperature majority rule involving species,
whose effect is to `convince' the central site to convert to
(or stay the same as) the types of its neighbours.
Next, outcomes for some process representing success or failure
are assigned to each of these sites in proportion to their type,
and last, based on these outcomes,
the central site could decide to switch to another type or stay the same.
This last step is rather subtle,
as it involves the comparison of two ratios,
$I^+_i(t)/N^+_i(t)$ and $I^-_i(t)/N^-_i(t)$,
so that for example even one individual who is successful
can have a greater impact on his/her neighbour
than say two individuals who are successful in a trio of the same type.
It is also in this last step that our interfacial
and cooperative models differ, in that the latter allows for long-term changes
even within a sea of the same species, depending on local outcomes.

This diversity of behavioural modelling is, however,
only one aspect of our work.
The resulting rich and novel collective behaviour is what,
for the most part, our paper is concerned with.
We have focused on the symmetric situation,
where none of the species is favoured.
In particular, the two-dimensional case of our interfacial model
turned out to be a physically motivated analogue of the kinetic Ising model
discussed in a more abstract fashion in the literature~\cite{O,DG}.
In three dimensions, the coarsening behaviour of the same interfacial model
has been characterised by a novel
dynamical critical exponent $\Omega\approx0.10$.
The latter exponent can be expected to describe the relaxation
of the energy of generic critical three-dimensional kinetic Ising models,
whose rules involve only surface noise.
The exponent $\Omega$ certainly deserves
to receive more attention in the future, in either three or higher dimensions.
In the cooperative model, our phase diagram involves,
in addition to regions of more conventional behaviour,
a phase whose coarsening behaviour we term `oscillatory'.
This is the first example to our knowledge of such a non-stationary
coarsening phenomenon,
where domain walls (whose interiors flash between one species and another)
appear to `breathe',
contracting and expanding forever in a more or less regular way.

We now come back to the effect of the dimensionality $d$
on the phase diagram of both our models.
This question is far from being trivial,
as some steps of the dynamics are rather subtle,
so that their net effect is hard to predict a priori.
We can, however, make some educated guesses about
the behaviour of our models.
Starting with the interfacial model,
we see that its one-dimensional version is frozen
for any value of the parameters $p$ and $\eps$.
Indeed, the dynamics of step~1 are equivalent to the zero-temperature
Glauber dynamics for the ferromagnetic Ising chain,
while the only effect of the other steps
is to reinforce the intensity of surface noise,
and the mobility of the kinks between the growing ordered domains.
This prediction is corroborated by numerical simulations,
which clearly show a fall-off of the mean energy
of the form $\mean{E}\approx A(p)t^{-1/2}$,
again because of an underlying diffusive coarsening behaviour~\cite{AB}.
The amplitude $A(p)$ is inversely proportional
to the mobility of the point defects discussed above.
We indeed find that $A(p)$ is a maximum at
$A(0)\approx A(1)\approx(8\pi)^{-1/2}=0.1995$
(the exact result for the Glauber-Ising chain),
while its minimum lies around $A(0.5)\approx0.1751$.
As the dimensionality increases,
the intermediate disordered phase is expected to open up as soon as $d>1$,
to get larger and larger
(we have indeed $p_{c2}-p_{c1}\approx0.14$ for $d=2$ and $0.41$ for $d=3$),
and to invade the whole phase diagram as $d\to\infty$.
This increase of the disordered phase with dimension
is not quantitatively reproduced by the pair approximation.
The latter approach is however far more powerful than
the standard mean-field theory,
which does not predict any sort of critical point in the interfacial model.
The situation of the cooperative model, with the hard rule of step~3a,
is opposite to that of the interfacial model,
and rather similar to that of standard equilibrium
statistical-mechanical models.
The one-dimensional model is always disordered,
as its dynamics is equivalent to that of the Ising chain at high temperature.
As dimension increases, the sizes $p_o$ of the oscillatory phase
and $1-p_c$ of the ferromagnetically ordered phase
are observed to increase slowly.

We now mention the many avenues we have left unexplored,
speculating on the likely behaviour of some of them,
based on our understanding of the phase diagram
of our models in the symmetric case.
First, we relegate to future investigations
a more detailed investigation of the interfaces in our interfacial model.
This issue has geometrical aspects,
such as the surface tension and its critical behaviour,
roughening phenomena, the cluster size distribution, and so on,
as well as purely dynamical aspects, such as persistence and ageing.
Another subject suitable for further development is the comparison
of various kinds of dynamics.
For a given model, i.e., given (stochastic) dynamical rules
to be applied at every site,
one can use either a parallel or a sequential updating procedure.
Sequential dynamics can be ordered
(the lattice is swept in a given ordered fashion),
disordered (lattice sites are chosen at random),
or more complex, such as the sublattice dynamics,
which consists in successively sweeping the even and the odd sublattices.
In the case of equilibrium statistical mechanical models,
these features affect neither the phase diagram nor critical properties.
In the present situation, changing the microscopic dynamical rules
(inside a reasonable class) could have a mild
but non-zero net effect on the phase diagram and on critical properties.

Although we have restricted the investigations presented in this paper
to the symmetric case when the performance parameters are identical,
(because our main interest in the problem
is largely to do with the critical behaviour at coexistence)
the model is really meaningful from a sociological point of view
only when $p_+>p_-$, with the possibility also of varying the convertibility
parameters $\eps_-$ and $\eps_+$.
In other words, we are led to consider the generic situation,
where the biasing fields $H$ and $B$, defined in eq.~(\ref{bias}),
are non-zero.
The case $p_+>p_-$ and $\eps_+>\eps_-$, i.e., $B<0<H$,
corresponds to the R species being more `genetically' successful,
but more fickle than the D species,
while for $p_+>p_-$ and $\eps_+<\eps_-$, i.e., both $H$ and $B$ positive,
the R species is both more `genetically' successful
and more retentive than the D species.
We would expect that in the latter case, the R species would usually
be globally more successful in the long term,
but that in the former case the final state of an initially random population
may depend on the ratio $H/\abs{B}$.
There would be rather interesting conclusions
from the point of view of societal evolution to be drawn
from these investigations.

Using the language of statistical mechanics,
each of the biasing fields $H$ and $B$ is equivalent to
a uniform magnetic field, favouring the R phase if it is positive.
We expect that there is a non-trivial coexistence line in the $(H,B)$ plane,
above which the R phase is favoured (the case of sociological interest),
and below which the D phase is favoured (its counterpart by symmetry).
The fate of the D phase, whenever it is unfavoured,
is dictated by the `equation of state' of the model,
relating the mean magnetisation $M$ to the applied fields $H$ and $B$.
In the regime where both fields are small,
the phase diagram of the symmetric situation investigated in this work
allows us to make the following qualitative predictions:
\begin{itemize}
\item{
In a disordered or paramagnetic phase, we have $M\approx\chi_1H+\chi_2B$
in the thermodynamic limit,
with both susceptibilities being finite and positive.
The mean population of the unfavoured species, $P=(1-M)/2$,
thus diminishes gradually,
in proportion to the intensity of the biasing fields.
In this regime, the coexistence line is linear: $H=(\chi_2/\chi_1)\abs{B}$.
}
\item{
In an ordered or ferromagnetic phase,
there is a non-zero spontaneous magnetisation $M_0$.
This means that the fraction of individuals in the unfavoured species,
which is $1/2$ in the symmetric case,
decreases discontinuously to $(1-M_0)/2$
whenever an `infinitesimal' biasing field is switched on.
For a large but finite sample of linear size $L$,
with $N\sim L^d$ individuals, fields of order $H\sim B\sim1/N$
are sufficient to bring on the decay of the unfavoured species.
The latter phenomenon is again driven by diffusion.
It thus lasts a time of order $\tau\sim L^2\sim N^{2/d}$.
}
\item{
In a totally ordered or frozen phase,
the spontaneous magnetisation assumes its maximal allowed value $M_0=1$.
In other words, all the individuals eventually belong to the favoured species,
in the presence of an `infinitesimal' biasing field.
The finite-size estimates of the previous point still hold true.
}
\item{
Less can be said {\em a priori} in the case where the symmetric model
is in the oscillatory phase, since this novel kind of order,
where the growth of clusters is modulated by intermittent conversions,
is not one where the usual phenomenology of equilibrium phase
transitions can be readily invoked.
We nevertheless speculate that the behaviour of mean populations
(possibly averaged over some intermediate time scale),
will be qualitatively similar to that observed in a ferromagnetically
ordered phase.
}
\end{itemize}

To close, we have presented in the above two models of competitive learning
that are respectively interfacial and cooperative in their behaviour.
We have focused on their properties in the symmetric situation
where neither of the species is favoured over the other one,
with emphasis on the phase diagram and on critical behaviour.
While many sociological consequences of our models remain to be explored
for the case when the species involved are unequally biased,
we believe that the behaviour at coexistence has already revealed a great deal
of the underlying physics.
The prediction of nontrivial critical behaviour
entirely driven by surface noise in our interfacial model,
or of the new phase of oscillatory coarsening
in the case of our cooperative model,
are possibly the most outstanding examples of this.
We believe that further investigations of the issues sketched in the discussion
could also yield phenomena, hitherto unknown, that are both novel in themselves
and completely natural to the evolution of such truly complex systems.

\subsubsection*{Acknowledgements}

AM gratefully acknowledges the hospitality of the Service de Physique
Th\'eorique, CEA Saclay, at which most of this work was carried out.
We are also pleased to thank Kalyan Chatterjee, J.M.~Drouffe, and C.~Godr\`eche
for interesting discussions.

\newpage
\section*{Table and caption}
{\vskip 1cm}
\begin{center}
\begin{tabular}{|c|c|c|c|}
\hline
& & &\\
model &\db{\mbox{transition points}}{\mbox{(pair approximation)}}
&\db{\mbox{transition points}}{\mbox{(numerical)}}
&\db{\mbox{characteristics}}{\mbox{of transitions}}\\
& & &\\
\hline
& & &\\
\db{\mbox{2D interfacial}}{\mbox{(square lattice)}}
&\dbl{p_{c1}=0.47598}{p_{c2}=0.83060}
&\dbl{p_{c1}=0.56\pm0.01}{p_{c2}=0.70\pm0.01}\Bigg\}
&\mbox{voter-like}\\
& & &\\
\hline
& & &\\
\db{\mbox{3D interfacial}}{\mbox{(cubic lattice)}}
&\dbl{p_{c1}=0.53047}{p_{c2}=0.86908}
&\dbl{p_{c1}=0.45\pm0.01}{p_{c2}=0.865\pm0.005}\Bigg\}
& $\Omega=0.10\pm0.04$\\
& & &\\
\hline
& & &\\
\db{\mbox{2D cooperative}}{\mbox{(square lattice)}}
&\tripl{\mbox{---}}{}{p_c=0.86805}
&\tripl{p_o=0.136\pm0.003}{}{p_c=0.873\pm0.002}
&\tripl{1/\nu_o=0.45\pm0.10}{}{\mbox{Ising-like}}\\
& & &\\
\hline
& & &\\
\db{\mbox{3D cooperative}}{\mbox{(cubic lattice)}}
&\tripl{\mbox{---}}{}{p_c=0.87319}
&\tripl{p_o=0.42\pm0.005}{}{p_c=0.82\pm0.01}
&\tripl{1/\nu_o=0.8\pm0.2}{}{\mbox{weakly Ising-like}}\\
& & &\\
\hline
\end{tabular}
\end{center}
{\vskip .6cm}
\noindent{\bf Table~1:}
Characterisation of phase transitions shown in Fig.~1.
Column~2: location of transition points within the pair approximation
(see Secs.~\ref{s2a} and~\ref{s3a}).
Columns~3 and 4: transition points and characteristics
(exponents, universality classes) of transitions,
obtained by means of numerical simulations
(see Secs.~\ref{s2n} and~\ref{s3n}).

\newpage
\section*{Figure captions}
{\vskip .6cm}

\noindent {\bf Figure~1:}
Qualitative phase diagram of interfacial and cooperative models
as a function of $p$.

\noindent {\bf Figure~2:}
Snapshots of the dynamics of the interfacial model on the square lattice,
for $p=0.7\approx p_{c2}$, at times (top) $t=8$, (bottom left) $t=64$,
and (bottom right) $t=512$.

\noindent {\bf Figure~3:}
Plot of the inverse energy $1/E(t)$ of the interfacial model
on the square lattice,
against $\ln t$, for various values of $p$, indicated on the curves,
in the vicinity of $p_{c2}\approx0.70$.
The dashed straight line has slope $2/\pi$ [see eq.~(\ref{ev})].

\noindent {\bf Figure~4:}
Log-log plot of the energy $E(t)$ of the interfacial model
on the cubic lattice,
against time $t$, for various values of $p$, indicated on the curves,
in the vicinity of $p_{c1}\approx0.45$.
The dashed lines, meant as guides to the eye,
have slopes 0 (disordered), $-1/2$ (ordered), and $-0.1$ (critical).

\noindent {\bf Figure~5:}
Log-log plot of the effective exponent $\Omega\eff(t)$ of the energy
of the interfacial model
on the cubic lattice, against time $t$, for the same values of $p$
as in Fig.~4.
The symbol to the right with an error bar corresponds to the estimate
$\Omega=0.10\pm0.04$ for the critical exponent.

\noindent {\bf Figure~6:}
Snapshots of the dynamics of the cooperative model on the square lattice,
for $p=0.1$, at times (top) $t=8$, (bottom left) $t=64$,
and (bottom right) $t=512$.

\noindent {\bf Figure~7:}
Snapshots of the dynamics of the cooperative model on the square lattice,
for $p=0.9$, at times (top) $t=8$, (bottom left) $t=64$,
and (bottom right) $t=512$.

\noindent {\bf Figure~8:}
Plot of the mean magnetisation $\mean{M}$
of the cooperative model on the square lattice,
against $p$, in the ferromagnetic phase.
The fit shown as a dashed line (see text) yields $p_c=0.873\pm0.002$.

\noindent {\bf Figure~9:}
Plot of the specific heat $C$
of the cooperative model on the cubic lattice
(multiplied by the transition probability $\Pi$), against $p$
near the ferromagnetic transition, for several sample sizes.

\noindent {\bf Figure~10:}
Plot of the position $p\max(L)$ of the peak in the data
of Fig.~9, against $L^{-1/\nu}$.
Dashed line: fit discussed in the text, incorporating a linear
correction to scaling.

\noindent {\bf Figure~11:}
Plot of the skew magnetisation $M\skew(t)$,
of the cooperative model on the square lattice,
against time $t$, for a single history of a sample of size~$100^2$,
with $p=0.1$.

\noindent {\bf Figure~12:}
Plot of the order parameter $\mean{M^2}$
of the cooperative model on the square lattice against $p$,
for several sample sizes.

\noindent {\bf Figure~13:}
Scaling plot of the data shown in Fig.~12,
illustrating the finite-size scaling law~(\ref{mfss})
at the transition between the oscillatory and disordered phases, with
$p_o=0.136$ and $1/\nu_o=0.45$.

\noindent {\bf Figure~14:}
Scaling plot of the order parameter $\mean{M^2}$
of the cooperative model on the cubic lattice,
illustrating the finite-size scaling law~(\ref{mfss})
at the oscillatory transition, with $p_o=0.422$ and $1/\nu_o=0.8$.


\newpage
\begin{thebibliography}{10}

\bibitem{L} T.M. Liggett, {\it Interacting Particle Systems} (Springer,
New-York, 1985).

\bibitem{O} M.J. de Oliveira, J.F.F. Mendes, and M.A. Santos, J. Phys.
{\bf A}: Math. Gen. {\bf 26}, 2317 (1993).

\bibitem{DG} J.M. Drouffe and C. Godr\`eche, J. Phys. {\bf A}: Math.
Gen. {\bf 32}, 249 (1999).

\bibitem{VM} J.T. Cox and D. Griffeath, Ann. Probab. {\bf 11}, 876
(1983); Contemp. Math. {\bf 41}, 55 (1985);
M. Bramson, J.T. Cox, and D. Griffeath, Probab. Th. Rel. Fields {\bf
77}, 613 (1988); J.T. Cox, Ann. Probab. {\bf 16}, 1559 (1988).

\bibitem{VP} M. Scheucher and H. Spohn, J. Stat. Phys. {\bf 53}, 279 (1988);
P.L. Krapivsky, Phys. Rev. A {\bf 45}, 1067 (1992);
L. Frachebourg and P.L. Krapivsky, Phys. Rev. E {\bf 53}, R~3009 (1996);
E. Ben-Naim, L. Frachebourg, and P.L. Krapivsky, Phys. Rev. E {\bf 53},
3078 (1996).

\bibitem{KI} R. Kikuchi, Phys. Rev. {\bf 81}, 988 (1951).

\bibitem{BU} D.M. Burley, in {\it Phase Transitions and Critical
Phenomena}, vol.~2 (Academic, London, 1972).

\bibitem{D} R. Dickman, Phys. Rev. A {\bf 34}, 4246 (1986).
For a review, see: J. Marro and R.~Dickman, {\it Nonequilibrium Phase
Transitions in Lattice Models}, Monographs and Texts in Statistical Physics
(Cambridge University Press, 1999).

\bibitem{AB} A.J. Bray, Adv. Phys. {\bf 43}, 357 (1994).

\bibitem{FSS} J. Cardy~(ed.), {\it Finite-Size Scaling} (North Holland,
Amsterdam, 1988); V. Privman~(ed.), {\it Finite-Size Scaling and Numerical
Simulation of Statistical Systems} (World Scientific, Singapore, 1990).

\bibitem{GUI} R. Guida and J. Zinn-Justin, J. Phys. {\bf A}: Math. Gen.
{\bf 31}, 8103 (1998).

\end{thebibliography}
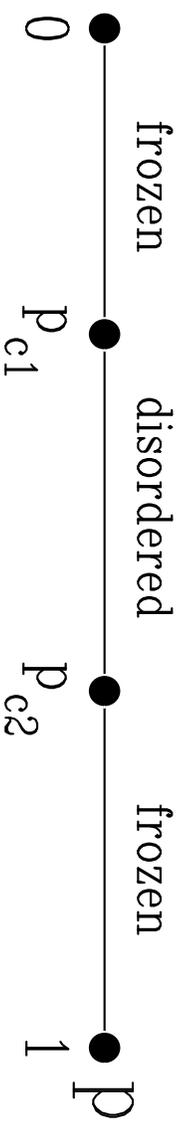
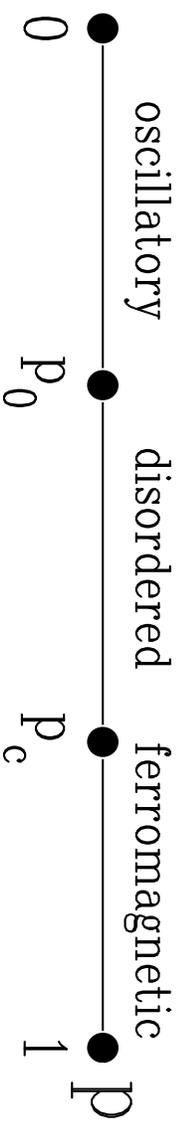
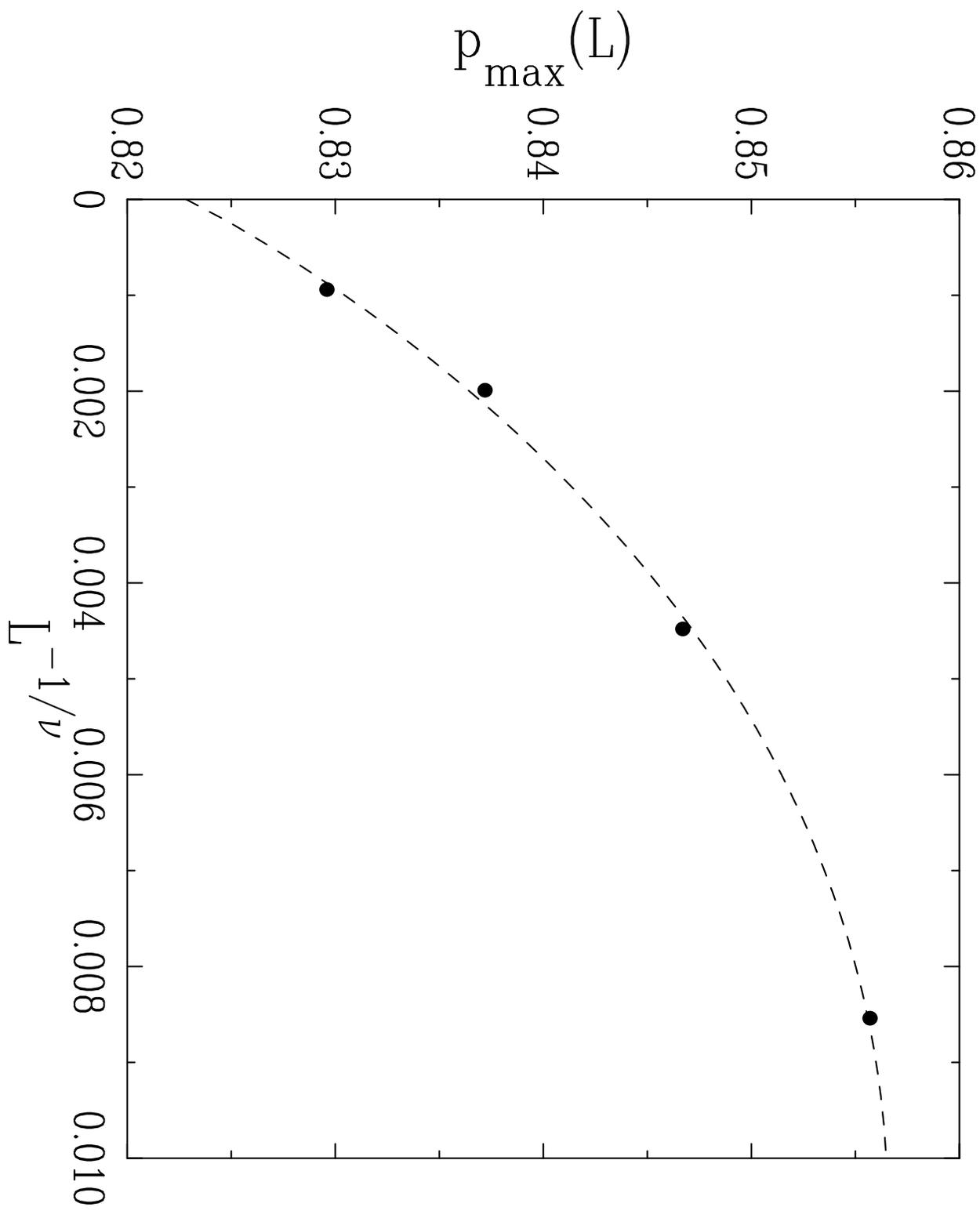
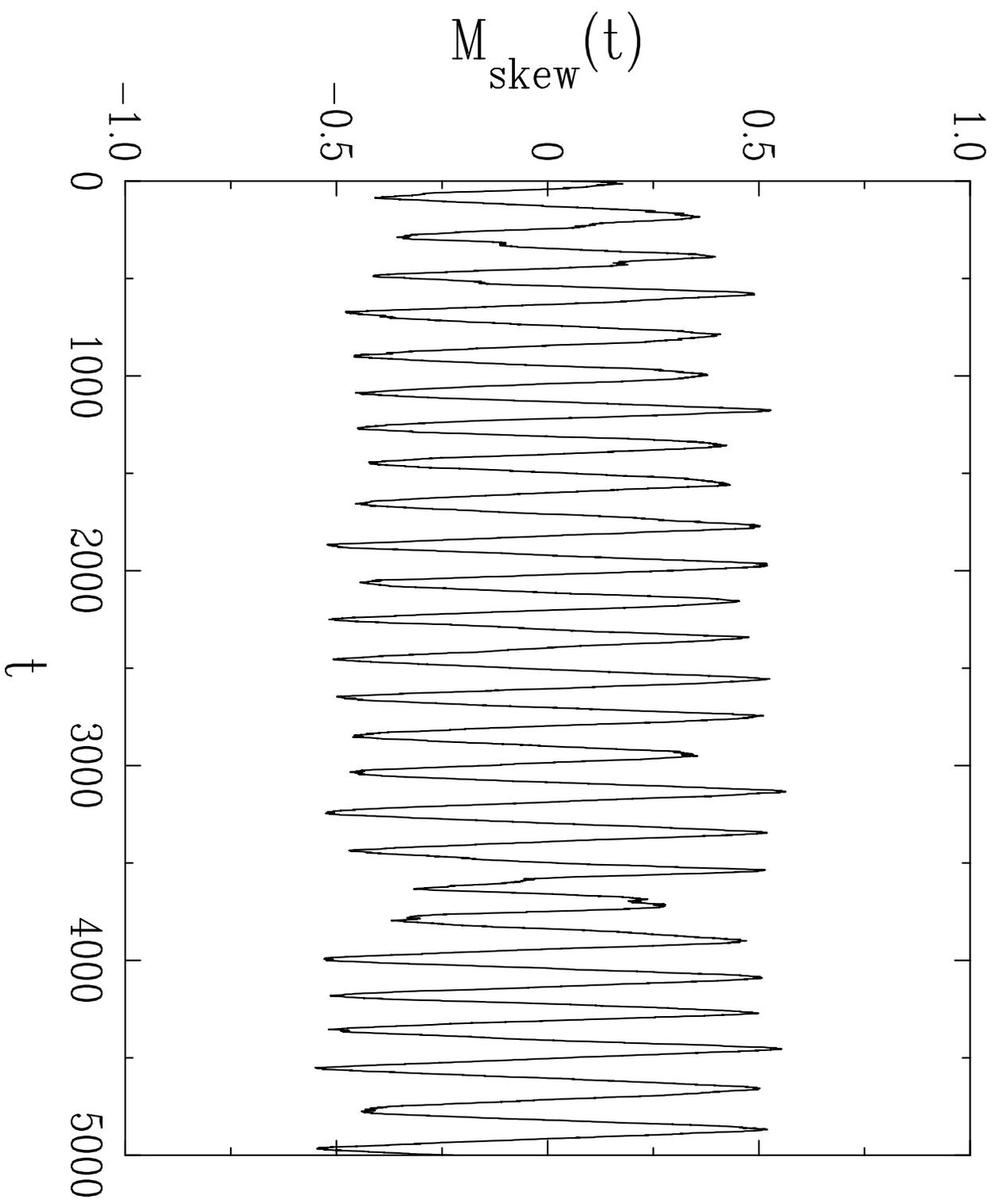
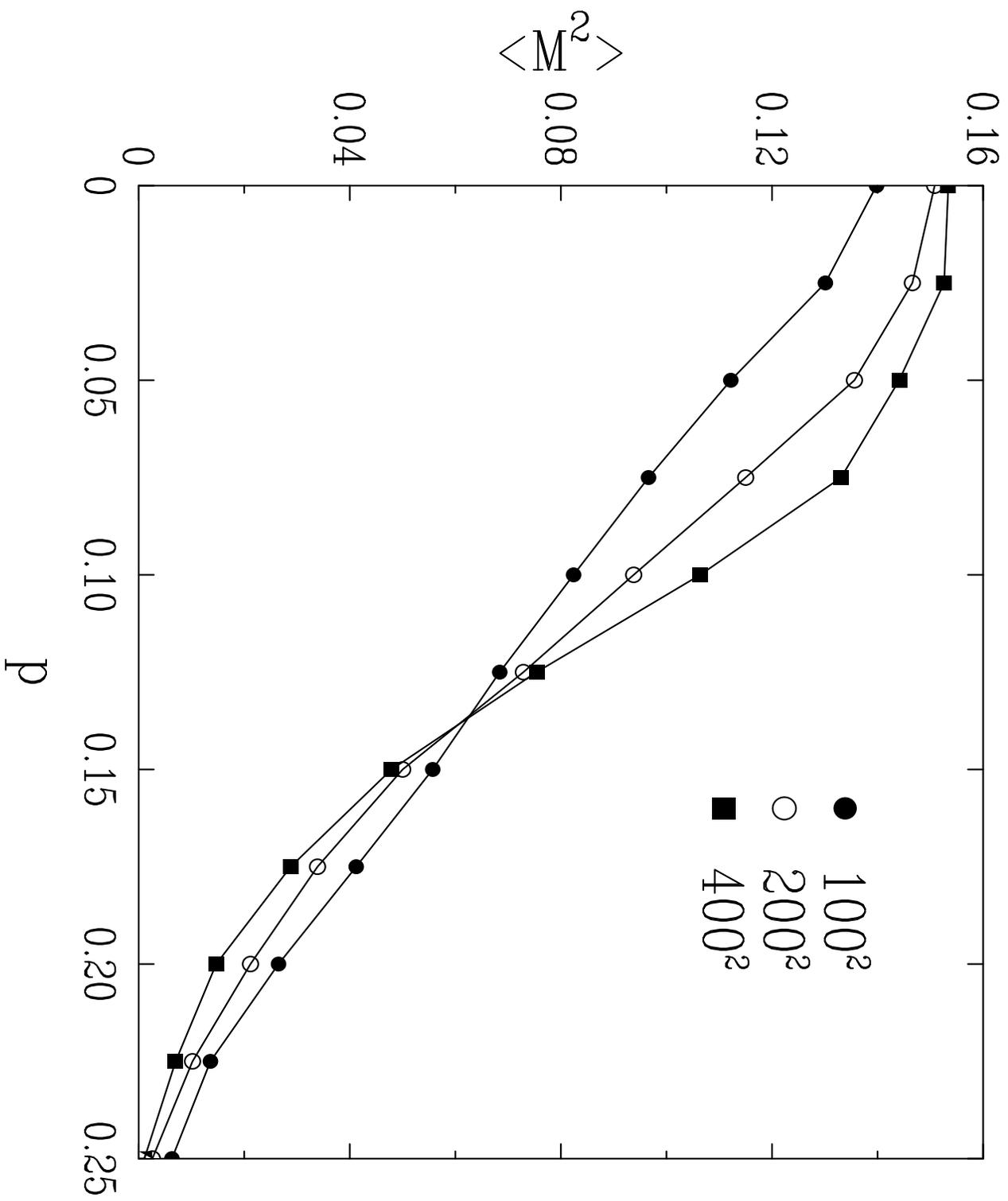
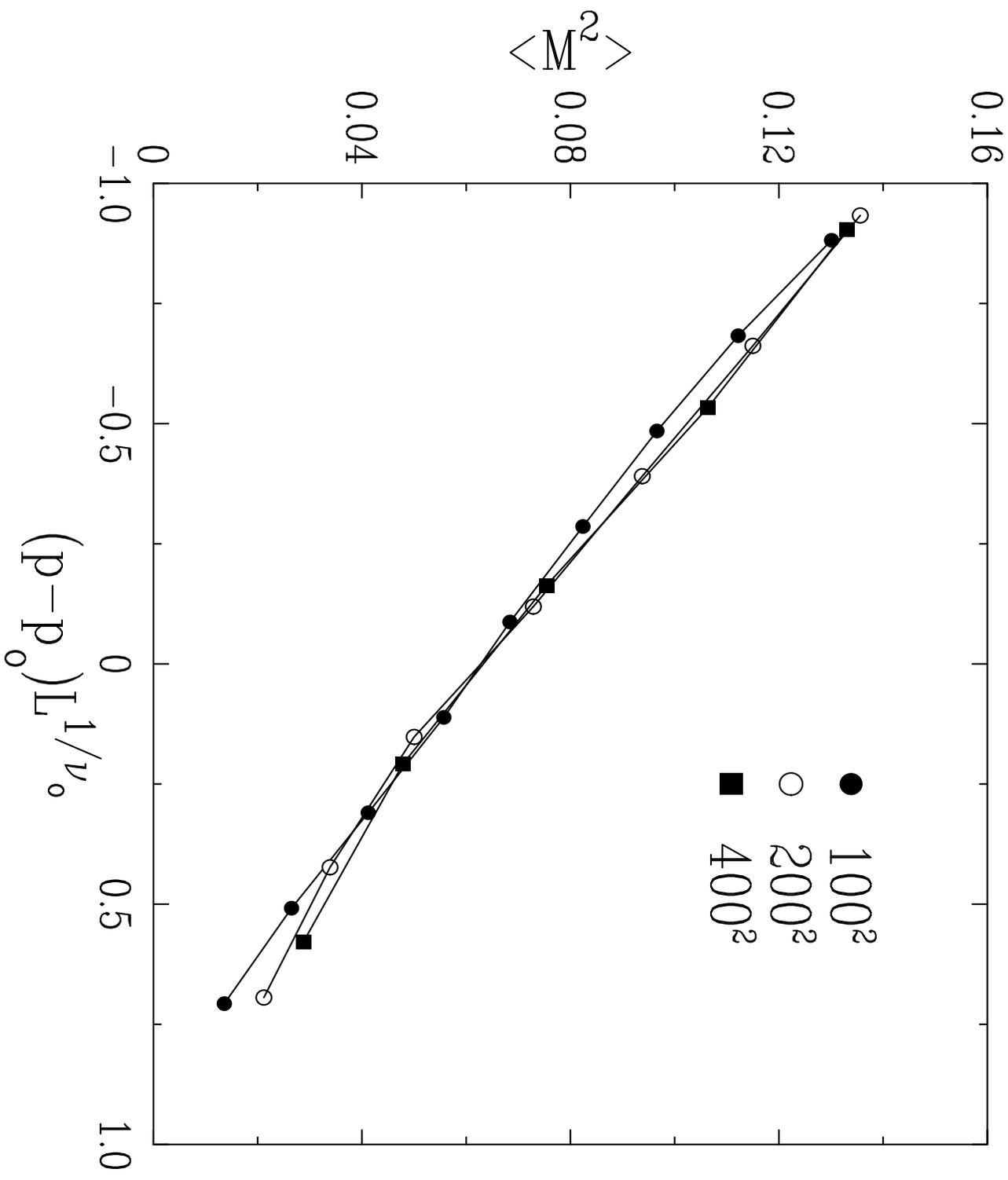
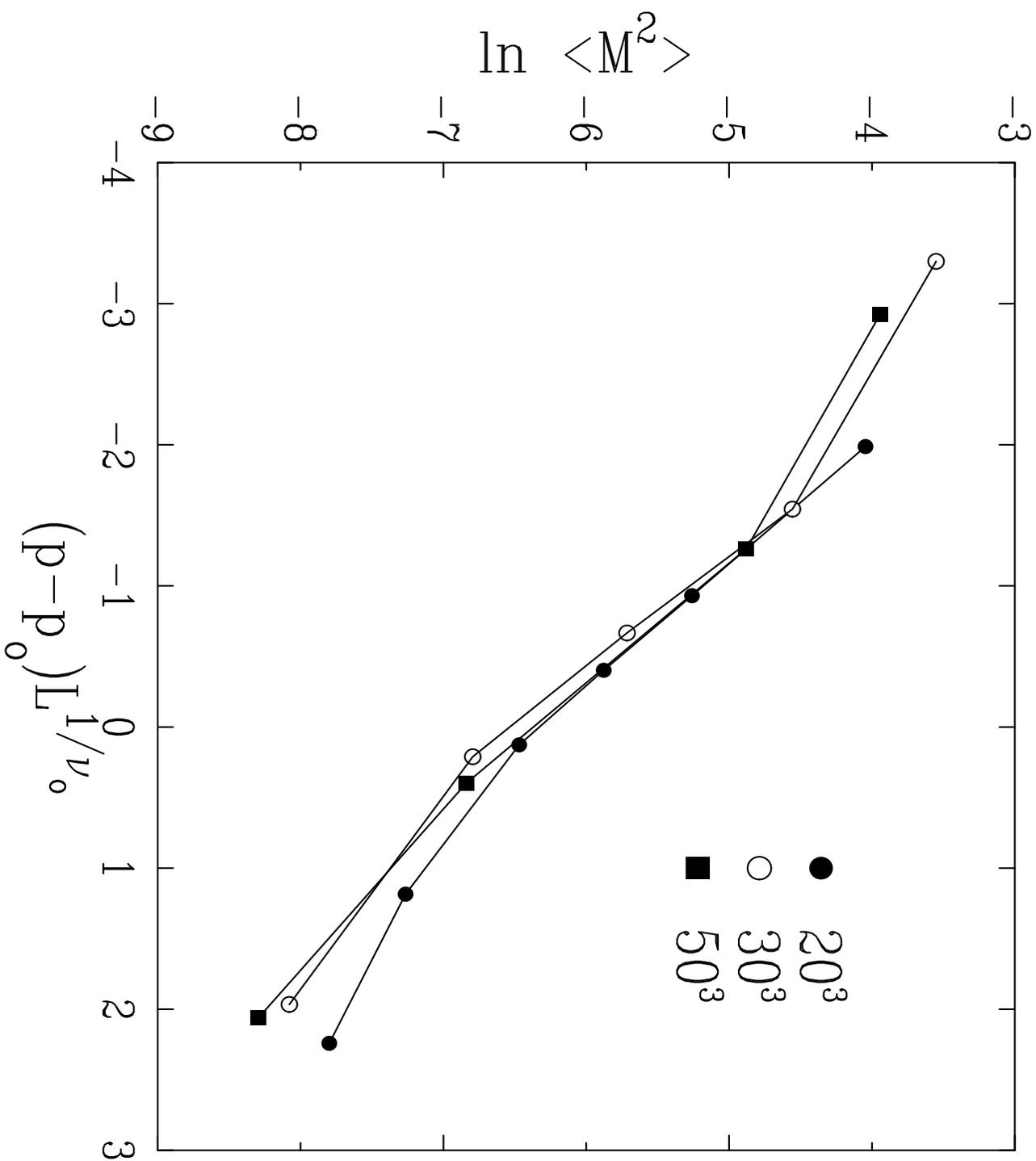
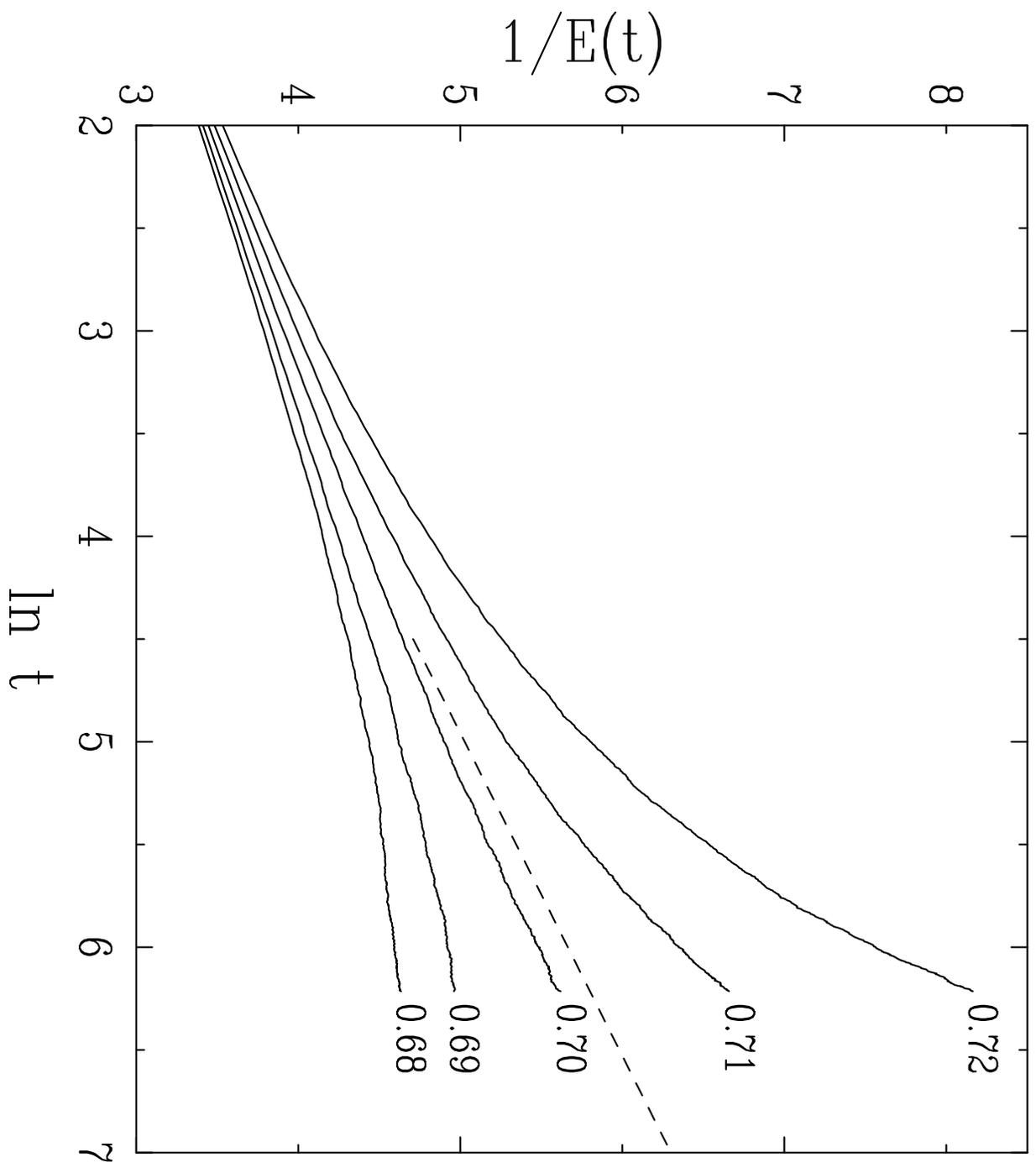
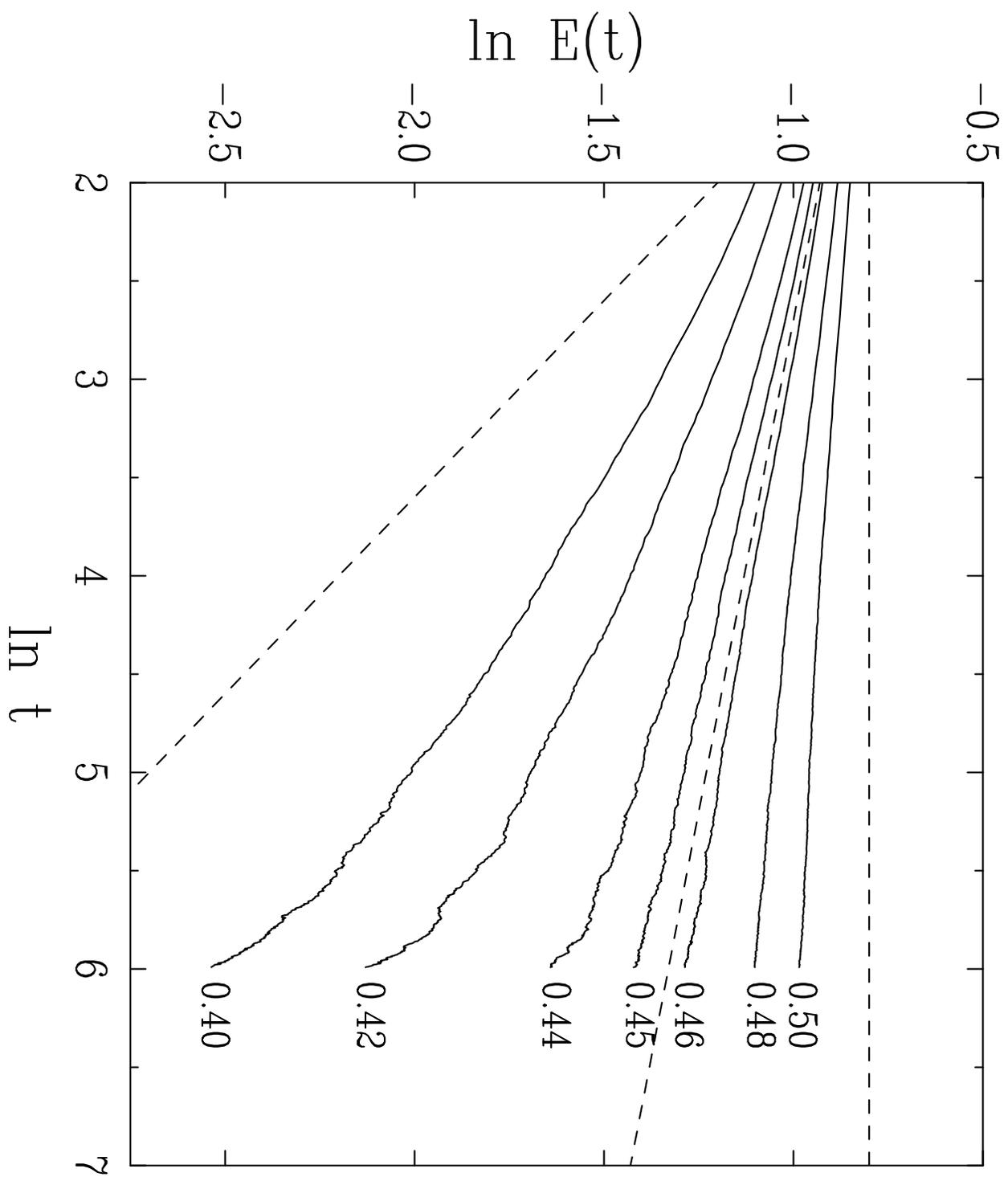
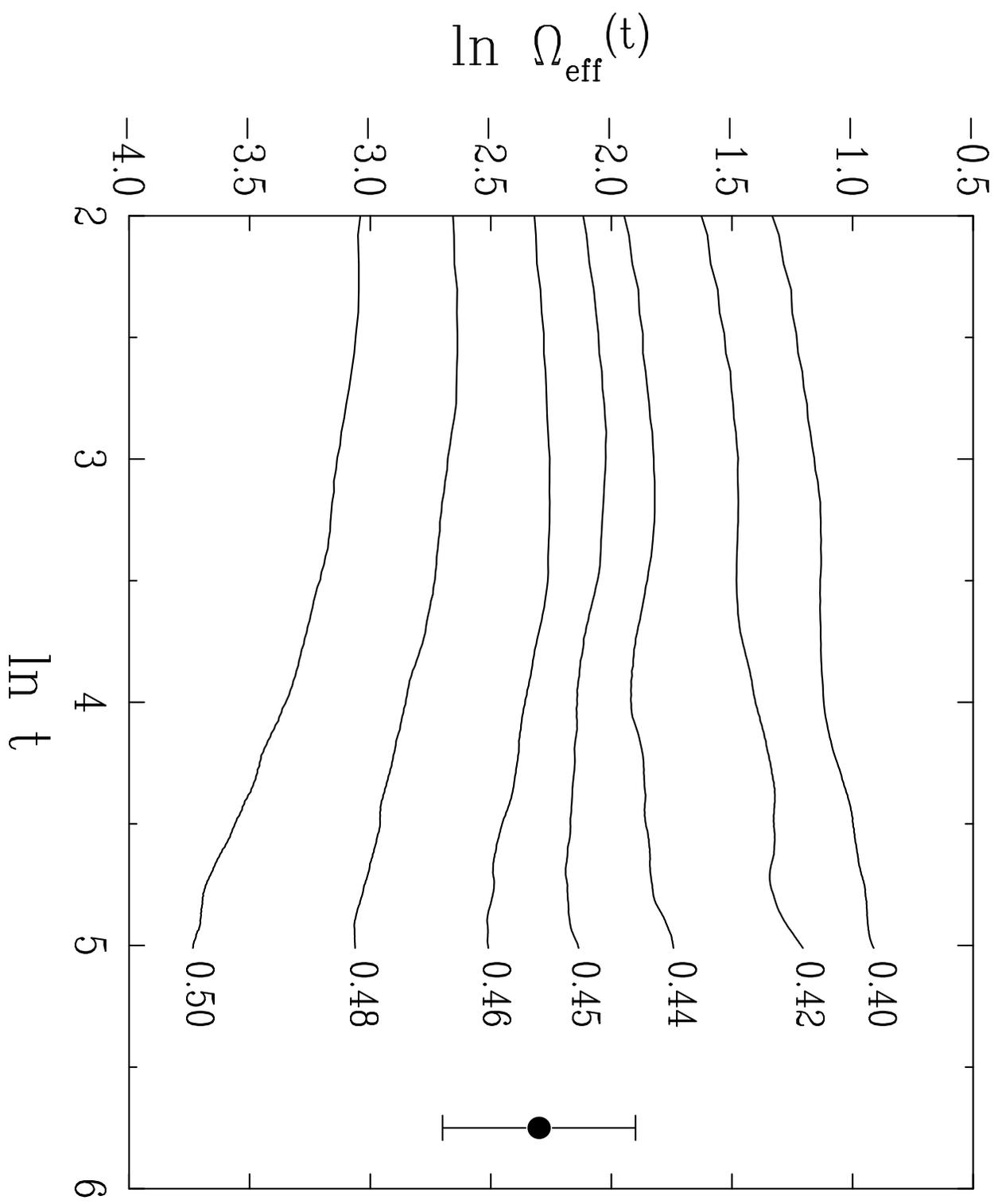
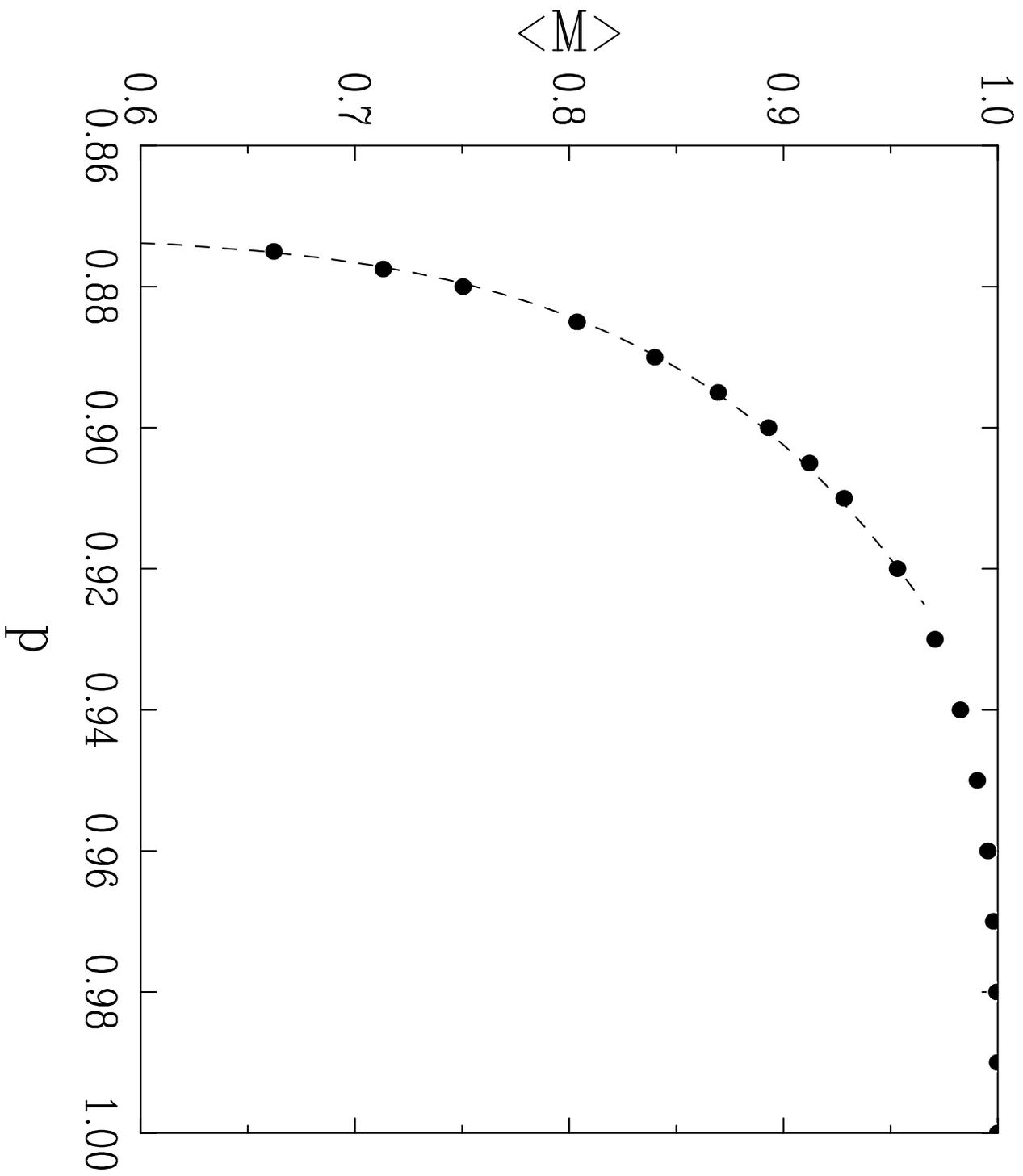
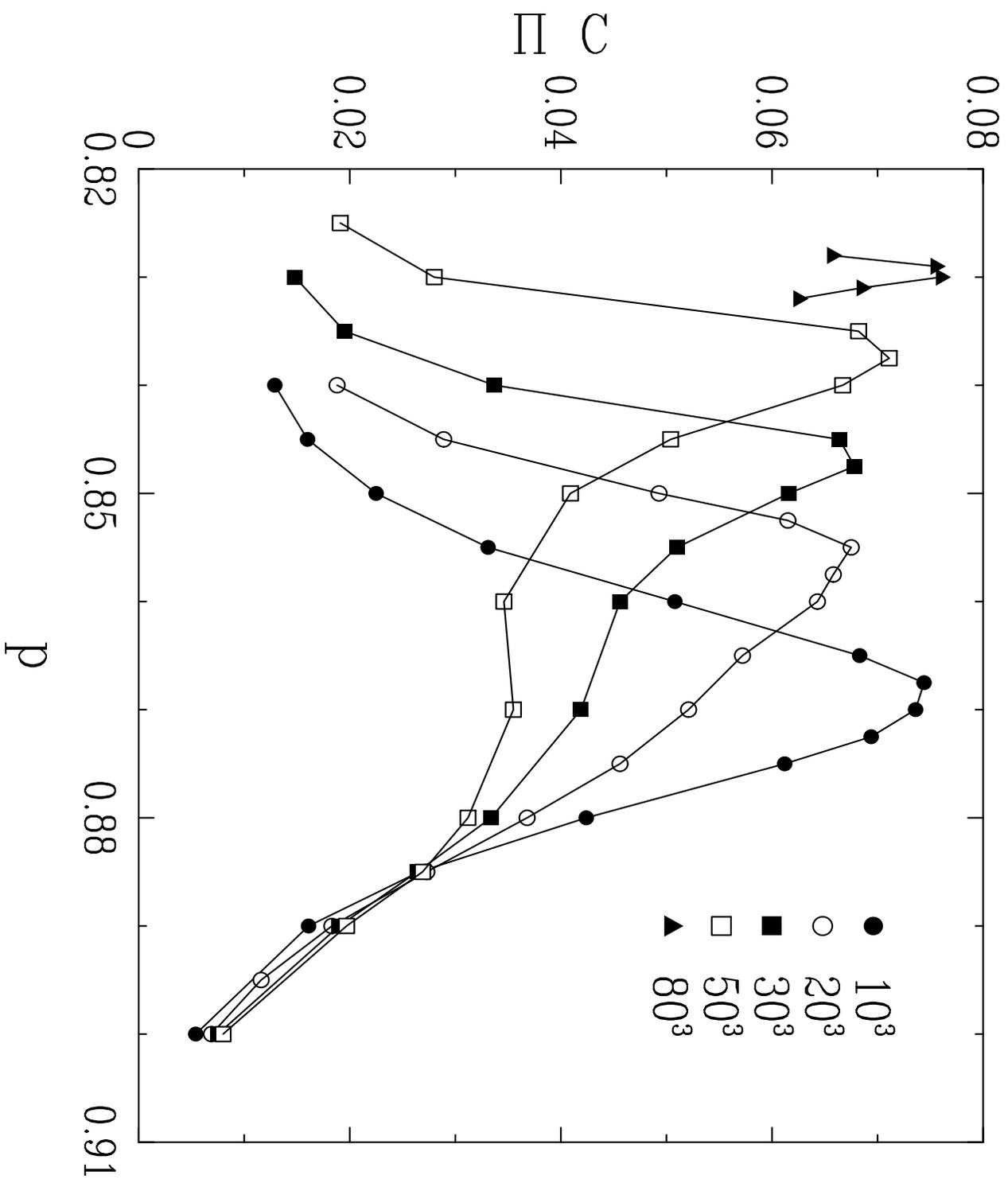
\end{document}